\documentclass[12pt, anon]{l4dc2024}

\newcommand{\lina}[1]{}
\newcommand{\yiwen}[1]{}
\newcommand{\zishuo}[1]{}
\newcommand{\revise}[1]{#1}

\usepackage{wrapfig}
\usepackage{caption}
\usepackage{afterpage}
\usepackage{booktabs}
\usepackage{diagbox}
\usepackage{makecell}
\usepackage{multirow}
\usepackage{siunitx}
\usepackage{longtable}

\usepackage{tikz}
\usepackage[utf8]{inputenc}
\usepackage{pgfplots}
\usetikzlibrary{shapes,arrows,fit,positioning,calc}
\usetikzlibrary{decorations.pathreplacing}
\usetikzlibrary{arrows.meta, decorations.pathmorphing}
\DeclareUnicodeCharacter{2212}{−}
\usepgfplotslibrary{groupplots,dateplot}
\usetikzlibrary{patterns,shapes.arrows}
\pgfplotsset{compat=newest}

\usepackage{bm}

\newcommand{\Rb}{{\mathbb{R}}}

\newcommand{\Sb}{{\mathbb{S}}}

\newcommand{\bL}{{\bm{L}}}

\DeclareMathOperator{\prox}{{Prox}}
\DeclareMathOperator{\proj}{{Proj}}

\DeclareMathOperator{\minimize}{{minimize}}
\DeclareMathOperator{\maximize}{{maximize}}

\newtheorem{problem}{Problem}

\title[MPC-Inspired Reinforcement Learning for Verifiable Model-Free Control]{
  MPC-Inspired Reinforcement Learning for Verifiable Model-Free Control
}
\usepackage{times}

\author{%
 \Name{Yiwen Lu} \Email{luyw20@mails.tsinghua.edu.cn}\\
 \Name{Zishuo Li} \Email{lizs19@mails.tsinghua.edu.cn}\\
 \Name{Yihan Zhou} \Email{zhouyh23@mails.tsinghua.edu.cn}\\
 \Name{Na Li} \Email{nali@seas.harvard.edu}\\
 \Name{Yilin Mo} \Email{ylmo@tsinghua.edu.cn}\\
}

\begin{document}

\maketitle

\vspace{-1.5cm}

\begin{abstract}%
  In this paper, we introduce a new class of parameterized controllers, drawing inspiration from Model Predictive Control (MPC). The controller resembles a Quadratic Programming (QP) solver of a linear MPC problem, with the parameters of the controller being trained via Deep Reinforcement Learning (DRL) rather than derived from system models. This approach addresses the limitations of common controllers with Multi-Layer Perceptron (MLP) or other general neural network architecture used in DRL, in terms of verifiability and performance guarantees, and the learned controllers possess verifiable properties like persistent feasibility and asymptotic stability akin to MPC. On the other hand, numerical examples illustrate that the proposed controller empirically matches MPC and MLP controllers in terms of control performance and has superior robustness against modeling uncertainty and noises. Furthermore, the proposed controller is significantly more computationally efficient compared to MPC and requires fewer parameters to learn than MLP controllers. Real-world experiments on vehicle drift maneuvering task demonstrate the potential of these controllers for robotics and other demanding control tasks.
\end{abstract}


\newcommand{\presectionspace}[1]{%
  \def\prespace{#1}
}

\newcommand{\postsectionspace}[1]{%
  \def\postspace{#1}
}

\newcommand{\mysection}[1]{%
  \vspace{\prespace}
  \section{#1}
  \vspace{\postspace}
}
\newcommand{\mysubsection}[1]{%
  \vspace{\prespace}
  \subsection{#1}
  \vspace{\postspace}
}

\presectionspace{-1em}
\postsectionspace{-0.25em}

\setlength{\floatsep}{5pt}
\setlength{\textfloatsep}{5pt}
\setlength{\intextsep}{5pt}

\setlength{\abovedisplayskip}{6pt plus 2pt minus 2pt}
\setlength{\belowdisplayskip}{6pt plus 2pt minus 2pt}
\setlength{\abovedisplayshortskip}{5pt plus 2pt minus 2pt}
\setlength{\belowdisplayshortskip}{5pt plus 2pt minus 2pt}

\mysection{Introduction}
\label{sec:intro}

Recent years have witnessed the development of Deep Reinforcement Learning (DRL) for control~\citep{lillicrap2015continuous,duan2016benchmarking,haarnoja2018learning}, with the locomotion of agile robots~\citep{xie2018feedback,li2021reinforcement,margolis2022rapid,rudin2022learning} being a notable example. Many such applications use Multi-Layer Perceptron~(MLP) as the entirety or a part of the control policy, which, despite their remarkable empirical performance, face limitations in terms of explainability~\citep{agogino2019challenges} and performance guarantees~\citep{osinenko2022reinforcement}. Research efforts have been devoted to the stability verification of MLP controllers~\citep{dai2021lyapunov,zhou2022neural}, structured controller parameterizations~\citep{srouji2018structured,johannink2019residual,sattar2020quickly,ni2021recurrent}, or a combination of both~\citep{zinage2023neural}, and the learning of explainable and verifiable controllers has remained an active topic.

\revise{
On the other hand, Model Predictive Control (MPC) has been a prevalent choice for high performance controller, and its stability and safety has been well-studied~\citep{morari1999model,qin2003survey,schwenzer2021review}. Recently, there is a growing interest in augmenting MPC with learning, a large portion of which are focused on addressing the challenges in designing critical components of MPC like prediction models~\citep{desaraju2016experience,soloperto2018learning,hewing2019cautious}, terminal costs and constraints~\citep{brunner2015stabilizing,rosolia2017learning,abdufattokhov2021learning}, stage costs~\citep{englert2017inverse,menner2019constrained}, or a combination of these components~\citep{gros2019data}; readers are referred to~\cite{hewing2020learning} for a comprehensive overview. Most of the methods follow a model-based framework -- using data to estimate a model and then perform optimal control in a receding horizon fashion. These methods still suffer various challenges, such as requiring intensive computation at each time horizon, and making myopic decisions that lead to infeasibility or inefficiency in the long run.
}


\revise{
Motivated by the advantages of DRL and MPC, we propose an MPC-inspired yet model-free controller.
Leveraging the fact that linear MPC solves a Quadratic Programming (QP) problem at each time step, we consider a parameterized class of controllers with QP structure similar to MPC. However, the key distinction lies in the approach to obtaining the QP problem parameters: instead of deriving them from a model, they are optimized via DRL. This approach ensures that the resulting controllers not only have theoretical guarantees akin to MPC, thanks to its QP structure, but also demonstrate competitive performance and computational efficiency when empirically compared to MPC and MLP controllers.
}

Contrasting with works from the learning community, \revise{such as \cite{amos2018differentiable}, which uses MPC as a module in a larger policy network, and \cite{ha2018world,hansen2023td,lecun2022path}, which adapt MPC ideas by planning in latent spaces}\lina{what's broader learning frameworks? either making it more specific or provide short examples?}\yiwen{Added brief description.}, our work retains the QP structure of linear MPC.
While these learning-based approaches aim to be general and tackle more challenging tasks by integrating MPC concepts into comprehensive models, they often include black-box components without control-theoretic guarantees. In contrast, our approach specializes in control tasks, emphasizing performance guarantees and computational efficiency. However, empirical evidence on a real-world robotic system demonstrates that our controller may generalize beyond simple linear systems, as illustrated in an aggressive vehicle control setting.


\textbf{Our Contribution.} In this paper, we propose a new parameterized class of MPC-inspired controllers.
\revise{Specifically, our controller resembles an unrolled QP solver, structured similarly to a Recurrent Neural Network~(RNN), with its parameters learned rather than derived via a predictive model.}
To train the parameters of the controller, most of the existing DRL methods, such as PPO~\citep{schulman2017proximal}, could be used.
However, in contrast to most DRL-trained controllers, which often lack rigorous theoretical guarantees, our MPC-inspired controller is proven to enjoy verifiable properties like persistent feasibility and asymptotic stability.
We also compare the proposed controller on benchmark tasks with other methods such as classical MPC and \revise{DRL-trained neural network controllers,}
showing that our proposed controller enjoys lighter computation and increased robustness.
Lastly, though we only provide theoretical guarantees for controlling a linear system, the generalizability of the proposed constroller is empirically demonstrated via vehicle drift maneuvering, a challenging nonlinear robotics control task, indicating potential applications of our controller to real-world nonlinear robotic systems.

\mysection{Problem Formulation and Preliminaries}
\label{sec:prelim}
\textit{Notations.} Subscripts denote the time index, e.g., $x_k$ stands for the system state at step $k$, and $x_{0:k}$ means the sequence $x_0, x_1, \ldots, x_k$.
Superscripts denote the iteration index in an iterative algorithm, e.g. $y^i$ stands for the variable $y$ at the $i$-th iteration. Bracketed subscripts denote slicing operation on a vector, e.g., $v_{[1]}$ denotes the first element of the vector $v$, and $v_{[1:n]}$ denotes its first $n$ elements. The set of positive definite $n\times n$ matrices is denoted as $\mathbb{S}^n_{++}$, and the nonnegative orthant of $\mathbb{R}^n$ is denoted as $\mathbb{R}^n_+$. The Kronecker product of two matrices $A$ and $B$ is denoted as $A \otimes B$. The block diagonal matrix with diagonal blocks $A_1, \ldots, A_n$ is denoted as $\mathrm{diag}(A_1, \ldots, A_n)$. \revise{The projection operator to a convex set $C$ is denoted as $\Pi_C(\cdot)$.}

\mysubsection{Problem Formulation}

In this paper, we consider the discrete-time infinite-horizon constrained linear-quadratic optimal control problem, formulated as follows:
\begin{problem}[Infinite-horizon constrained linear-quadratic optimal control]
\label{prob:infinite_horizon}
\begin{subequations}
\begin{align}
\underset{u_{0:\infty}}{\text{minimize}} &\quad \limsup_{N \to \infty} \frac1{N} \sum_{k=0}^{N-1} (x_{k+1} - r)^\top Q (x_{k+1} - r) + u_k^\top R u_k, \label{eq:cost} \\
 \text{subject to} &\quad x_{k+1} = A x_k + B u_k, \label{eq:nominal_sys} \\
&\quad u_{\min} \leq u_k\leq u_{\max}, x_{\min} \leq x_{k+1} \leq x_{\max}, \label{eq:constraints}
\end{align}
\end{subequations}
\lina{I would prefer changing the equation numbers to (1a), (1b), (1c)}
where $x_k \in \mathbb{R}^{n_{sys}}$ are state vectors, $u_k \in \mathbb{R}^{m_{sys}}$ are control input vectors, $r \in \mathbb{R}^{n_{sys}}$ is the reference signal, $A \in \mathbb{R}^{n_{sys} \times n_{sys}}$ and $B \in \mathbb{R}^{n_{sys} \times m_{sys}}$ are the system and input matrices, $Q \in \mathbb{S}^{n_{sys}}_{++}$ and $R \in \mathbb{S}^{m_{sys}}_{++}$ are the stage cost matrices, and $u_{\min}, u_{\max} \in \mathbb{R}^{m_{sys}}$ and $x_{\min}, x_{\max} \in \mathbb{R}^{n_{sys}}$ are bounds on control input and state respectively.
It is assumed without loss of generality that $(A, B)$ is controllable.
\end{problem}



\mysubsection{Linear MPC and its QP Representation}

Problem~\ref{prob:infinite_horizon} is typically computationally intractable due to infinite planning horizons and constraints. A commonly adopted approximation is to truncate it to finite horizon $N$, and solve the problem formulated in Problem~\ref{pb:MPC} at each time step, with $x_0$ being the current state. The first control input $u_0^*$ from the optimal solution is applied to the system in a receding horizon fashion.

\vspace{-0.25em}
\begin{problem}[Linear MPC]\label{pb:MPC}
    \begin{subequations}
    	\begin{align}
    		\underset{x_{1:N}, u_{0:N-1}}{\text{minimize}} &\quad  \sum_{k=0}^{N-1} (x_{k+1}-r)^\top Q (x_{k+1}-r) + u_k^\top R u_k, \\
    		 \text{subject to} \quad & \quad x_{k+1} = A x_{k} + B u_k, \quad k = 0, \cdots, {N-1}, \label{eq:constraints_MPC} \\
    		& \quad u_{\min} \leq u_k\leq u_{\max}, x_{\min} \leq x_{k+1} \leq x_{\max} , \quad k = 0, \cdots, {N-1}.
    	\end{align}
    \end{subequations}
 \lina{similarly, change the equation number be (2a),(2b),(2c)}\yiwen{Done.}
\end{problem}

The above MPC problem can be cast into a Quadratic Programming~(QP) problem in the following standard form\footnote{Although a linear MPC without terminal costs or constraints is presented here for simplicity, one can derive a similar QP formulation for linear MPC with quadratic terminal cost and affine terminal constraint.}:

\vspace{-0.25em}
\begin{problem}[Standard-form QP]\label{pb:QP}
  \begin{equation}
  \begin{aligned}
    \underset{y}{\text{minimize}} &\quad \frac{1}{2} y^{\top} P y+q^{\top} y, \quad \text{subject to} \quad H y+b\geq 0,
  \end{aligned}
  \label{eq:qp}
  \end{equation}
where $y \in \mathbb{R}^{n_{qp}}$, $P \in \mathbb{S}^{n_{qp}}_{++}$, $q \in \mathbb{R}^{n_{qp}}$, $H \in \mathbb{R}^{m_{qp} \times n_{qp}}$, and $b \in \mathbb{R}^{m_{qp}}$.
\end{problem}

The translation from Problem~\ref{pb:MPC} to Problem~\ref{pb:QP} can be performed by using the control sequence $y = \begin{bmatrix}
  u_0^\top &\cdots &u_{N-1}^\top
\end{bmatrix}^\top$ as the decision variable, and eliminating the equality constraints~\eqref{eq:constraints_MPC} by representing the trajectory $x_{1:N}$ using $y$. The resulting QP problem size and parameters are:
\begin{align}
  & n_{qp} = N m_{sys}, \quad m_{qp} = 2N (m_{sys} + n_{sys}), \\
  & P = \boldsymbol{B}^\top \boldsymbol Q \boldsymbol B +\boldsymbol R, \quad q = 2 \boldsymbol B^\top\boldsymbol Q (\boldsymbol A x_0 - \boldsymbol r), \quad
  H = -\boldsymbol C \boldsymbol B -\boldsymbol D, \quad b =\boldsymbol e -\boldsymbol C\boldsymbol A x_0,
  \label{eq:mpc2qp}
\end{align}
where
\begin{equation}
  \begin{footnotesize}
	\bm{A}= \begin{bmatrix}
		A \\
		A^2\\ \vdots \\ A^{N}
	\end{bmatrix},
	\bm{B}= \begin{bmatrix}
		B & &&&\\
		AB & B &&& \\
		\vdots&&\ddots&\\
		A^{N-1}B &\cdots & &B
	\end{bmatrix},
  \quad
  \begin{aligned}
  & \bm{C} = I_N \otimes \mathrm{diag}(I_{n_{sys}}, -I_{n_{sys}}, \mathbf{0}_{2m_{sys} \times 2m_{sys}}), \\
  & \bm{D} = I_N \otimes \mathrm{diag}(\mathbf{0}_{2n_{sys} \times 2n_{sys}}, I_{m_{sys}}, -I_{m_{sys}}), \\
  & \bm{e} = I_N \otimes [x_{\max}^\top \; -x_{\min}^\top \; u_{\max}^\top\; -u_{\min}^\top]^\top, \\
  & \bm{r}=I_N \otimes r, \bm{Q}= I_N \otimes Q , \bm{R}= I_N \otimes R.
  \end{aligned}
  \end{footnotesize}
  \label{eq:AB}
\end{equation}

\mysubsection{Algorithm for Solving QPs}
\label{sec:pdhg}


A family of efficient methods for solving QPs is operator splitting algorithms~\citep{ryu2022large}, which are adopted by existing solvers such as OSQP~\citep{stellato2020osqp}.
An iteration of an operator splitting algorithm for solving QPs can generally be represented as a combination of affine transformations and projections on the variable. For example, \revise{we derive a variant of the Primal-Dual Hybrid Gradient (PDHG)~\citep{chambolle2011first} algorithm, whose iteration can be expressed in the succinct form shown as follows:}
\footnote{\revise{
    The form~\eqref{eq:pdhg_iter} is slightly different from the original PDHG iteration~\citep[p. 75]{ryu2022large} in that the primal and dual variables are updated synchronously, making it conceptually more straightforward to draw similarity between solver iterations and neural networks; see~\cite[Appendix A]{lu2023bridging} for the details.
}}
\begin{equation}
  z^{i+1} = \Pi_{ \mathbb{R}^{m_{qp}}_+} \left( (I - 2\alpha F)z^i + \alpha(I - 2F)\lambda^i) - 2\alpha \mu \right), \quad \lambda^{i+1} = F(z^i + \lambda^i) + \mu,
  \label{eq:pdhg_iter}
\end{equation}
where $z = Hy + b \in \mathbb{R}^{m_{qp}}$ is the primal variable of an equivalent form of the original problem~\eqref{eq:qp}, $\lambda \in \mathbb{R}^{m_{qp}}$ is a dual variable introduced by the same equivalent form, $\alpha > 0$ is the step size, and the parameters in the iteration are:
\begin{equation}
  F = (I + HP^{-1}H^\top)^{-1}, \quad \mu = F(HP^{-1}q - b).
  \label{eq:pdhg_affine_param}
\end{equation}
Once one obtains an approximate solution $z^i$, the original variable can be recovered from the equality-constrained QP problem $y^i \in \operatorname{arg\,min} \{ \frac12 y^\top P y + q^\top y \mid Hy + b = z^i \}$, where $y^i$ can be explicitly represented as:
\begin{equation}
  y^i = -P^{-1} q + P^{-1}H^\top (HP^{-1}H^\top)^\dagger (z^i - b + HP^{-1}q).
  \label{eq:get_sol}
\end{equation}

\vspace{-0.25em}
\begin{theorem}
  If $0 < \alpha < 1$ and the problem~\eqref{eq:qp} is feasible, then the iterations~\eqref{eq:pdhg_iter} yields $y^i \to y^*$,
  where $y^i$ is in~\eqref{eq:get_sol} and $y^*$ is the optimal solution of the original problem. Furthermore, the suboptimality gap satisfies:
  \begin{equation}
    p^i - p^* \leq \| \lambda^i \|_2 \| r_{prim}^i \|_2 + \| y^i - y^* \|_2 \| r_{dual}^i \|_2,
  \end{equation}
  where $p^i, p^*$ are the primal value at iteration $i$ and the optimal primal value respectively, and $r_{prim}^i, r_{dual}^i$ are the primal and dual residuals defined as follows:
  \begin{equation}
    r_{prim}^i = Hy^i + b - z^i, r_{dual}^i = Py^i + q + H^\top \lambda^i.
  \end{equation}
  \label{thm:pdhg}
\end{theorem}

\vspace{-2em}
\begin{proof}
\revise{Results similar to Theorem~\ref{thm:pdhg} has been derived in \citep{chambolle2011first}\citep{boyd2011distributed}. Therefore, the full proof is presented in the extended version of the paper~\cite[Appendix A]{lu2023bridging} due to space limit.}
\end{proof}
\yiwen{The iteration~\eqref{eq:pdhg_iter} is a variant of the original algorithm with lower memory footprint. The theorem is a collection of separate results in the literature (convergence of solution + suboptimality bound). They are not directly available from literature, but our contribution is essentially some algebra and restatements. I added a footnote on this.}\lina{I think it is important to explain it right after the Theorem 1 saying briefly that Theorem 1 is similar to *** in \cite{}. Because iteration ~\eqref{eq:pdhg_iter} is a variant of ... restate what you replied. It is important for readers to know the real contribution we want to present, not wasting time on theorems like this and for some reviewers, they are very picky if authors claim contributions which are not }

The iteration~\eqref{eq:pdhg_iter} on the primal-dual variable pair \revise{$(z, \lambda)$} can be implemented by interleaving an affine transformationwhose parameters $(F, \mu)$ depend on the problem parameters $(P, q, H, b)$, and a projection of the \revise{$z$-part} onto the positive orthant, which is equivalent to ReLU activation in neural networks.
Therefore, the sequence of iterations for solving a QP problem resembles a single-layer \revise{Recurrent Neural Network (RNN)} with weights dependent on the QP parameters $(P, q, H, b)$, followed by ReLU activation. This resemblance facilitates the end-to-end policy gradient-based reinforcement learning of the QP parameters, as is discussed in Section~\ref{sec:method}.

\vspace{1em}

\mysection{Learning Model-Free QP Controllers}
\label{sec:method}


This section introduces a reinforcement learning framework to tune the parameters of QP problem~\eqref{eq:qp}, instead of deriving them using the predictive models via conventional MPC. This could be beneficial to combat the short-sightedness~\citep{erez2012infinite} and the lack of robustness~\citep{forbes2015model}, which may occur in MPC. 

\begin{figure}
  \centering

\begin{tikzpicture}[node distance=1cm,auto,>=latex']
    \tikzstyle{learnable block} = [draw, thick, fill=blue!20, rectangle, minimum height=3em, minimum width=3em]
    \tikzstyle{block} = [draw, thick, rectangle, minimum height=3em, minimum width=3em]
    \tikzstyle{layer} = [block, minimum height=2em]
    \tikzstyle{input} = []
    \tikzstyle{sum} = [draw, fill=blue!20, circle, node distance=1.5cm]
    \tikzstyle{decision} = [diamond, draw, text width=5.2em, aspect=1.6, text centered, inner sep=0pt]
    \tikzstyle{line} = [draw, -latex']

    \tikzset{
        my split node/.style={
            block,
            rectangle split,
            rectangle split parts=2,
            rectangle split draw splits=false, 
            inner sep=2.9pt, 
            append after command={
              \pgfextra{
                \draw (\tikzlastnode.north west) -- (\tikzlastnode.north east);
              }
            }
          },
          rectangle split part fill={blue!20,white}
    }

    \tikzset{
        invisible/.style={
            draw=none,
            fill=none,
            text opacity=0
        }
    }

    \node [input, name=input1] {$x_0$};
    \node [input, below=0.2cm of input1.south, anchor=north] (input2) {$r$};
    \node [learnable block, right=of input1, xshift=-0.5cm, yshift=-0.25cm, text width=2.6em] (affine1) {$W_q$ \\ $W_b, b_b$};
    \node [my split node,
    right=1.25cm of affine1, anchor=center] (combined)
    { $P,H$ \nodepart{two} $q,b$ };
    \draw (combined.west) -- (combined.east);
    \node [block, invisible, right=1.25cm of affine1, node distance=3cm, anchor=north, minimum height=1.5em] (qb) {$q,b$};
    \node [layer, right=2.2cm of combined.east] (affine2) {\small Affine};
    \node [layer, below=0.5cm of affine2] (projection) {\small ReLU};
    \node [layer, right=0.5cm of affine2] (affine3) {\small Affine};
    \node [layer, right=0.5cm of projection] (projection3) {\small ReLU};
    \node [layer, right=1.cm of affine3] (affine4) {\small Affine};
    \node [layer, right=1.cm of projection3] (projection4) {\small ReLU};

    \node [learnable block, below=0.5cm of combined, draw=none, xshift=-0.8cm, minimum height=2em] (learnable) {Learnable Parameters $\theta$};

    \path [line] (input1) -- (affine1.west |- input1);
    \path [line] (input2) -- (affine1.west |- input2);
    \path [line] (affine1.east |- qb) -- node[pos=0.35] {\eqref{eq:qb_affine}} ($(qb.west)!0.5!(qb.center)$);
    \coordinate (iter top) at ($(affine2.north) + (0, 0.4)$);
    \path [line] (combined.east) --node [midway] {\eqref{eq:pdhg_affine_param}} ++(0.7, 0) |-  (iter top)  -- (affine2.north);
    \path [line] ($(affine2.west) + (-1.2, 0)$) -- node [pos=0.5, below, text width=4em] {$(z^0, \lambda^0)$ \\ $=(0, 0)$} (affine2.west);
    \path [line] (affine2) -- (projection);
    \path [line] (projection.east) -| ($(affine2)!0.5!(affine3)$) |- (affine3.west);
    \path [line] (affine3) -- (projection3);
    \path [line] (projection3.east) -- ($(projection3.east)!0.25!(projection4.west)$) |- ($(affine3)!0.5!(affine4)$);
    \path [line] ($(projection3.east)!0.5!(projection4.west)$) -| ($(affine3.east)!0.75!(affine4.west)$) -- (affine4.west);
    \path [line] (affine4) -- (projection4);
    \node [right=1cm of affine4.east] (output) {$u_0=y^{n_{iter}}_{[1:m_{sys}]}$};
    \path [line] (projection4.east) -- ++(0.25, 0) |- node [pos=0.1, right] {$(z^{n_{iter}}, \lambda^{n_{iter}})$} node [near end] {\eqref{eq:get_sol}} (output.west);
    \node at ($(projection3.east)!0.5!(affine4.west)$) {$\ddots$} ;
    \path [line] (iter top -| affine3.north) -| (affine4.north);
    \path [line] (iter top) -| (affine3.north);

    \draw [decorate,decoration={brace,amplitude=5pt,mirror,raise=1ex}]
    (projection.south) -- (projection4.south) node[midway,below=1.5ex] (repeat text){Repeat $n_{\text{iter}}$ times};

    \coordinate (box left) at ($(affine2.west) + (-0.1, 0)$);
    \coordinate (box right) at ($(affine4.east) + (0.1, 0)$);
    \coordinate (box top) at ($(iter top) + (0, -0.1)$);
    \coordinate (box bottom) at ($(repeat text.south) + (0, -0.05)$);
    \node(pdhgbox)[fit=(box left) (box right) (box top) (box bottom),
    draw=green!50!black,
    dashed,
    thick,
    inner xsep=0.0cm,
    inner ysep=0.0cm,
    label={[text=green!50!black]below:QP solver iterations}]
    {};


\end{tikzpicture}
  \caption{\small Proposed control policy architecture. The controller solves a QP problem in form~\eqref{eq:qp}, whose parameters $P, H$ are shared across all initial state and reference $(x_0, r)$, while $q, b$ depend affinely on $(x_0, r)$ with weights $W_q, W_b$ and bias $b_b$ (see~\eqref{eq:qb_affine}). An approximate solution to the QP problem, $y^{n_{iter}}$, is obtained by running a $n_{iter}$ QP solver iterations~\eqref{eq:pdhg_iter} followed by a transform~\eqref{eq:get_sol}, whose first $m_{sys}$ dimensions are used as the current control input $u_0$.
  }
  \label{fig:policy_architecture}
\end{figure}
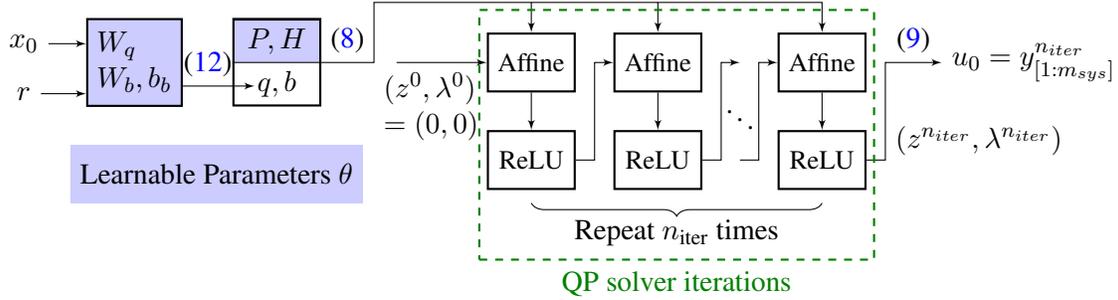

The policy architecture facilitating this learning process is shown in Figure~\ref{fig:policy_architecture}, where the control policy, i.e., the mapping from $x_0$ and $r$ to the control action $u_0$, is represented as a fixed number of PDHG iteration for solving a QP problem, or equivalently, an unrolled RNN, the parameters of which can then be trained using most of the existing policy-gradient based or actor-critic DRL method.

\lina{have we explained RNN yet? if not, we need to explain.}\yiwen{I expanded the above two sentences a bit, and pointed out the connection to the previous section. Do they read clear?}
Here we highlight several critical design components of the policy architecture. The first two are regarding the parameterization of the policy, i.e., \emph{what} to learn:

\noindent$\bullet$ \textbf{State-independent matrices $P, H$:} Note from~\eqref{eq:mpc2qp} that for the MPC controller, the matrices $P, H$ holds the same across different initial and reference state $(x_0, r)$'s. Motivated by this fact, the matrices $P, H$ are state-independent in the proposed policy architecture, i.e., only one matrix $P$ and one matrix $H$ need to be learned for a specific system. Additionally, to ensure the positive definiteness of $P$, we use the factor $L_P$ in the Cholesky decomposition $P = L_P L_P^\top$ instead of the matrix $P$ itself as the learnable parameter, and force its diagonal elements to be positive via a softplus activation~\citep{zheng2015improving}, a commonly applied trick for learning positive definite matrices~\citep{haarnoja2016backprop,lutter2019deep}.

\noindent$\bullet$ \textbf{Affine transformations yielding vectors $q, b$:} \revise{Continuing with the inspirations from MPC}~\eqref{eq:mpc2qp}, we restrict the vectors $q$ to depend linearly on the current state $x_0$ and the reference state $r$, and the vector $b$ to depend affinely on $x_0$:
\begin{equation}
  q(x_0, r; W_q) = W_q [x_0^\top\; r^\top]^\top, \quad b(x_0, r; W_b, b_b) = W_b x_0 + b_b,
  \label{eq:qb_affine}
\end{equation}
where $W_q, W_b$ (resp. $b_b$) are learnable matrices (resp. vector) of proper dimensions.

\vspace{10pt}
The above-described parameterization strategy ensures that when the chosen problem dimensions $n_{qp}, m_{qp}$ match the dimensions of the QP translated from MPC, then the MPC policy is within the family of parameterized policies defined by the proposed architecture. In other words, the proposed controller can be viewed as a generalization of MPC. While the state-independence and state-affineness constraints in the parameterization limit the range of policies relative to those in MLP or MPC-like policies with generic function approximator components~\citep{amos2018differentiable}
, this reduced complexity is beneficial for deriving provable theoretical guarantees, as is discussed in Section~\ref{sec:theory}. On the other hand, both numerical examples and real-world experiments show that the empirical performance of the proposed controller matches that of an MPC or MLP controller, and thus is not hindered by this restriction.
\lina{Yiwen, I revised this sentence. I think you might want to take a look to see how we can change the sentence in a more positive tone.}\yiwen{Understood! It reads better when we emphasize the benefits.}

Another two design components determine \emph{how} the parameters are learned:

\noindent$\bullet$ \textbf{Unrolling with a fixed number of iterations:} To solve the QP problem and differentiate the solution with respect to the problem parameters, we deploy a fixed number $n_{iter}$ of QP solver iterations described in Section~\ref{sec:pdhg}, and differentiate through the computational path of these iterations, a practice known as unrolling~\citep{monga2021algorithm}. Unlike implicit differentiation methods~\citep{amos2017optnet,amos2018differentiable,agrawal2019differentiable}, which differentiate through the optimality condition and hence requires the forward pass of the solver to reach the stationary point, our method directly differentiates the solution after $n_{iter}$ iterations, and can obtain a correct gradient even if the stationary point is not reached within these iterations. According to our empirical results, a small number of iterations would suffice for good control performance (e.g., $n_{iter} = 10$), which mitigates the computational burden of the unrolling process. Intuitively, the sufficiency of a small $n_{iter}$ can be accredited to the model-free nature of the proposed method, which, by discarding the restrictions imposed by model-based prediction (see~\eqref{eq:AB}), gains the flexibility to learn a QP problem that not only optimizes the controller performance, but also is easy to solve.

\noindent$\bullet$ \textbf{Reinforcement learning with residual minimization:}
The control policy described above, parameterized by $\theta = (L_P, H, W_q, W_b, b_b)$, can serve as a drop-in replacement for standard policy networks, and be optimized using various off-the-shelf policy-based or actor-critic RL algorithms, such as PPO~\citep{schulman2017proximal}, SAC~\citep{haarnoja2018learning} and DDPG~\citep{lillicrap2015continuous}.
However, apart from the standard RL loss, we also include a regularization term for minimizing the residuals given by the QP solver embedded in the policy. Given a dataset $\mathcal{D}$ of transition samples, it is defined as follows:
\begin{equation}
  \ell_{res}(\theta; \mathcal{D}) = \frac{1}{| \mathcal{D} |} \sum_{k = 1}^{| \mathcal{D} |} \| H y^{n_{iter}}_k + b_k - z_k^{n_{iter}} \|_2^2 + \| P y^{n_{iter}}_k + q_k + H^\top \lambda^{n_{iter}}_k \|_2^2,
  \label{eq:residual}
\end{equation}
which, motivated by the result stated in Theorem~\ref{thm:pdhg} that small residuals are indicative of near-optimality, encourages the learned QP problems to be easy to solve.
From above, the procedure of policy learning using an RL algorithm is shown in Algorithm~\ref{alg:main}.

\begin{algorithm2e}
  \caption{Framework of Learning of QP Controllers}
  \label{alg:main}
  \DontPrintSemicolon
  \LinesNumbered
  \KwIn{Simulation environment $Env$ with nominal dynamics~\eqref{eq:nominal_sys}, RL algorithm $RL$, policy architecture $\pi_\theta$ shown in Fig.~\ref{fig:policy_architecture}, regularization coefficient $\rho_{res}$}
  \KwOut{Optimized policy parameters $\theta = (L_P, H, W_q, W_b, b_b)$}
  \For{epoch $= 1,2,\dots$}{
    Interact with $Env$ using current policy $\pi_\theta$ to collect a dataset $\mathcal{D}$\;
    Compute RL loss, denoted by $\ell_{RL}(\theta; \mathcal{D})$\;
    Compute residual loss $\ell_{res}(\theta; \mathcal{D})$ using~\eqref{eq:residual}\;
    Update $\theta$ according to the loss $\ell_{RL}(\theta; \mathcal{D}) + \rho_{res} \ell_{res}(\theta; \mathcal{D})$\;
  }
\end{algorithm2e}

As an additional note, the learned QP problem parameterized by $(P, H, q, b)$ can be enforced to be feasible, which facilitates the theoretical analysis of the learned controller. Readers are referred to \cite[Appendix B]{lu2023bridging} for the details.

\mysection{Performance Guarantees of Learned QP Controller}
\label{sec:theory}
\lina{after each theorem, try to add, The proof is available in ...\cite{ouronlinereport}}
In this section, we propose a method for establishing performance guarantees of a learned QP controller with the architecture described in Section~\ref{sec:method}. We provide sufficient conditions for persistent feasibility and asymptotic stability of the closed-loop system under a QP controller, which parallel the theoretical guarantees for linear MPC~\citep{borrelli2017predictive}. For simplicity, we consider the stabilization around the origin, i.e., $r = 0$, but the method of analysis can be extended to the general case. Additionally, we assume throughout the section that the optimal solution of the learned QP problem is attained, which can be ensured by allowing the QP solver to run sufficient iterations until convergence when deploying.

Denote the property under consideration as $\mathcal{P}$. Suppose that a certificate to $\mathcal{P}$, given the initial state $x_0$ is in a polytopic set $\mathcal{X}_0$, can be written in the following form:
\begin{equation}
  \min_{x_0\in \mathcal{X}_0, u_0, \nu} \left\{ f(x_0, u_0, \nu) | g(x_0, u_0, \nu) \leq 0, u_0 = \pi_\theta(x_0) \right\} \geq 0 \Rightarrow \mathcal{P}\text{ holds when } x_0 \in \mathcal{X}_0,
  \label{eq:certificate}
\end{equation}
where $\pi_\theta$ denotes the $\theta$-parameterized control policy described in Section~\ref{sec:method}, $\nu$ is an auxiliary variable, and $f, g$ are quadratic (possibly nonconvex) functions. The optimization problem in the LHS of~\eqref{eq:certificate} can be expressed as a bilevel problem by explicitly expanding the control policy $\pi_\theta$ as:
\begin{align}
  & \pi_\theta(x_0) = y^*_{[1:m_{sys}]}, y^* \in \operatorname{arg\,min} \left\{ (1/2) y^\top P y + q^\top y \mid Hy + b \geq 0 \right\}, \nonumber\\
  & \text{where } q = W_q x_0, b = W_b x_0 + b_b.
  \label{eq:qp_opt}
\end{align}
Replacing the inner-level problem in~\eqref{eq:qp_opt} by its KKT condition, the verification problem in~\eqref{eq:certificate} can be cast into a nonconvex Quadratically Constrained Quadratic Program (QCQP) with variables $x_0, \nu, y, \mu$.
Various computationally tractable methods for lower bounding the optimal value of a QCQP are available, such as Lagrangian relaxation~\citep{d2003relaxations} and the method of moments~\citep{lasserre2001global}, and once a nonnegative lower bound is obtained, the property $\mathcal{P}$ is verified.

Verification of persistent feasibility and asymptotic stability both fall into the framework described above. The conclusions are stated as follows:

\vspace{-0.25em}
\begin{theorem}[Certificate for Persistent Feasibility]
  The control policy~\eqref{eq:qp_opt} if persistently feasible (i.e., gives a valid control input that keeps the next state inside the bounds at every step) for all initial states $x_0 \in \mathcal{X}_0 = \{ x | Gx \leq c \}$, if the optimal value of the following nonconvex QCQP is nonnegative:
  \begin{align*}
  \underset{x_0, \nu, y, \mu}{\text{minimize}} & \quad -\nu^\top(G(Ax_0+By_{[1:m_{sys}]}) - c), \\
  \text{subject to} & \quad Gx_0 \leq c,  \nu \geq 0,  \mathbf{1}^\top \nu = 1, \\
  & \quad Py + W_q x_0  - H^\top \mu = 0,
    Hy + W_b x_0 + b_b \geq 0, \mu \geq 0,
    \mu^\top(Hy + W_b x_0 + b_b) = 0.
  \end{align*}
  \label{thm:feasibility}
\end{theorem}

\vspace{-2em}

To certify asymptotic stability, we consider the Lyapunov function of a stabilizing baseline MPC, and attempt to show that the Lyapunov function decreases along all trajectories even if the learned QP controller is deployed instead of the baseline MPC. A similar technique has been applied to the stability analysis of approximate MPC~\citep{schwan2023stability}. To formalize this idea, we define the following notations: $l(x, u) = x^\top Q x + u^\top R u$ is the stage cost; the baseline MPC policy has horizon $N$, terminal constraint $x_{N} \in \mathcal{X}_f$ and terminal cost $V_f(x_N)$; the function $J(x_0, u_{0:N-1}) = \sum_{k = 0}^{N-1} l(x_{k}, u_k) + V_N(x_N)$, where $x_{k+1} = Ax_k + Bu_k$, is the objective function of the baseline MPC. To ensure that the baseline MPC is stabilizing as long as it is feasible, one can choose $\mathcal{X}_f$ to be an invariant set under a stabilizing linear feedback controller $u = Kx$, and $V_f(x)$ to be the cost-to-go under $u = Kx$. Based on these notations, a certificate for asymptotic stability can be stated as follows:

\vspace{-1em}
\begin{theorem}[Certificate for Asymptotic Stability]
  Let $\mathcal{X}_0 = \{ x | Gx \leq c \}$ be a set where the baseline MPC is well-defined and the policy~\eqref{eq:qp_opt} is persistently feasible.
  The closed-loop system under \eqref{eq:qp_opt} is asymptotically stable on $\mathcal{X}_0$, if $b_b \geq 0$, and there exists $\epsilon > 0$ and $N \in \mathbb{N}^*$, such that the optimal value of the following problem is nonnegative:
  \begin{align*}
  \underset{x_0, \bar{u}_{0:N}, y, \mu}{\text{minimize}} & \quad J(x_0, \bar{u}_{0:N}) + l(x_0, y_{[1:m_{sys}]}) - J(x_0, (y_{[1:m_{sys}]}, \bar{u}_{1:N})) - \epsilon \| x_0 \|^2, \\
  \text{subject to} & \quad Gx_0 \leq c,
   x_{N+1}(Ax_0+By_{[1:m_{sys}]}, \bar{u}_{1:N}) \in \mathcal{X}_f, \\
  & \quad Py + W_q x_0  - H^\top \mu = 0,
    Hy + W_b x_0 + b_b \geq 0, \mu \geq 0,
    \mu^\top(Hy + W_b x_0 + b_b) = 0.
  \end{align*}
  \label{thm:stability}
\end{theorem}

\vspace{-2em}

\revise{
Proofs of Theorems~\ref{thm:feasibility} and~\ref{thm:stability}, as well as numerical examples showcasing the verification of a learned controller on a double integrator system, are provided in~\cite[Appendix D]{lu2023bridging}.
}


\mysection{Benchmarking Results}
\label{sec:benchmark}

In our empirical evaluations, we aim to answer the following questions:
\vspace{-0.5em}
\begin{itemize}
  \itemsep-0.25em 
  \parsep0em 
  \item How does the learned QP controller compare with common baselines (MPC, RL-trained MLP) on nominal linear systems?
  \item Can the learned QP controller handle modeling inaccuracies and disturbances?
  \item Does the method generalize to real-world robot systems with modeling inaccuracy and nonlinearity?
\end{itemize}
\vspace{-0.5em}
We only briefly describe the experimental setup and typical results in this section, with complete details on systems, setup, hyperparameters, baseline definitions, and additional results in~\cite[Appendix E]{lu2023bridging}.
Code is available at \url{https://github.com/yiwenlu66/learning-qp}.

\mysubsection{Results on Nominal Systems}
\label{sec:benchmark_nominal}

We compare the Learned QP (LQP) controller with MPC and MLP baselines on benchmark systems like the quadruple tank~\citep{johansson2000quadruple} and cartpole~\citep{geva1993cartpole}, generating random initial states and references across $10^4$ trials.
For MPC, we evaluate variants with and without manually tuned terminal costs over short (2 steps) and long (16 steps) horizons, all implemented using OSQP~\citep{stellato2020osqp}, a solver known for its efficiency in MPC applications~\citep{forgione2020efficient}, with default solver configurations.
Both LQP and MLP are trained using PPO, maintaining consistent reward definitions and RL hyperparameters.
We incrementally increase the MLP size until further increases yield negligible performance improvements, selecting this size for comparison.
The LQP is assessed in both small (\(n_{qp}=4, m_{qp}=24\)) and large (\(n_{qp}=16, m_{qp}=96\)) configurations, approximately aligning with the QP problem sizes from short- and long-horizon MPC. All training (including simulation and policy update) are performed on a single NVIDIA RTX 4090 GPU, with the small and large configurations taking 1.2 hours and 2.7 hours respectively.

\begin{table}[!htbp]
  \caption{\small Performance comparison on benchmark systems.}\label{tab:common}
  \small
  \vspace{-10pt}\setlength\tabcolsep{3pt} 
  \renewcommand{\arraystretch}{1.0} 
  \begin{tabular}{>{\centering\arraybackslash}m{2cm}|ccccc|ccccc}
    \toprule
    \multirow{2}{*}{\diagbox[dir=NW,innerwidth=2cm,height=2\line]{Method}{Metrics}} & \multicolumn{5}{c|}{Quadruple Tank} & \multicolumn{5}{c}{Cartpole Balancing} \\
    \cline{2-11}
    & Fail\% & Cost & P-Cost & FLOPs & \#Params & Fail\% & Cost & P-Cost & FLOPs & \#Params \\
    \midrule
    MPC(2)  & 16.59 & 236.1 & 275.7 & 95K$_{+ \text{1.2M}}$ & - & 100.0 & 1.36 & 129 & 67K$_{+ \text{814K}}$ & - \\
    MPC(16) & 4.27 & 228.3 & 237.2 & 22M$_{+ \text{52M}}$ & - & 46.86 & \underline{0.34} & 8.39 & 3.9M$_{+ \text{47M}}$ & - \\
    MPC-T(2) & 4.23 & 239.6 & 248.5 & 470K$_{+ \text{779K}}$ & -  & 100.0 & 1.41 & 122 & 89K$_{+ \text{792K}}$ & - \\
    MPC-T(16) & 3.22 & \textbf{224.8} & \underline{231.5} & 26M$_{+ \text{49M}}$ & - & 4.74 & \textbf{0.30} & \textbf{0.79} & 51M$_{+ \text{50M}}$ & - \\
    RL-MLP & \textbf{0.03} & 266.7 & 266.7 & \underline{23K} & 11K & \textbf{3.23} & 0.57 & 0.91 & \underline{87K} & 43K \\
    \textbf{LQP(4, 24)} & 0.18 & 272.5 & 272.8 & \textbf{14K} & \textbf{0.3K} & \underline{3.49} & 0.76 & 1.12 & \textbf{14K} & \textbf{0.2K} \\
    \textbf{LQP(16, 96)} & \underline{0.13} & \underline{227.3} & \textbf{227.6} & 208K & \underline{2.6K} & 4.11 & 0.44 & \underline{0.87} & 208K & \underline{2.2K} \\
    \bottomrule
  \end{tabular}
  \vspace{0.25em}

  {
    \footnotesize
  Methods: MPC($N$) = MPC (Problem~\ref{pb:MPC}) with horizon $N$ without terminal cost; MPC-T($N$) = MPC with horizon $N$ and manually tuned terminal cost; RL-MLP = reinforcement learning controller with MLP policy; LQP($n_{qp},m_{qp}$) = proposed learned QP controller with problem dimensions $(n_{qp},m_{qp})$. Metrics: Fail\% = percentage of early-terminated trials due to constraint violation; Cost = average LQ cost until termination; P-Cost = average cost with penalty for constraint violation; FLOPs = floating point operations per control step (reported as $\text{median}_{+(\text{max} - \text{median})}$ for variable data); \#Params = number of learnable policy parameters. Best is highlighted in \textbf{bold}, and second best is \underline{underlined}.
  }
\end{table}

The results of the benchmarking experiments are summarized in Table~\ref{tab:common}. In terms of control performance, LQP demonstrates comparable effectiveness to both MPC and MLP baselines. A benefit of LQP is its independence from manual tuning of the terminal cost, which can be important for MPC methods. Regarding computational efficiency, LQP stands out for its minimal demand for achieving similar control performance. This efficiency stems from LQP's fixed number of unrolled QP solver iterations. While MPC's computation cost varies based on implementation, the light computation of LQP is still noteworthy, especially in scenarios with tight computational limits. For example, the LQP(4, 24) configuration, despite having lowest FLOPs among all methods, still manages acceptable control performance. Finally, in terms of the number of learnable policy parameters, LQP requires substantially fewer than the RL-MLP. This hints at LQP's suitability for memory-constrained embedded systems and applicability to online few-shot learning.

Results on additional systems, including a numerical example of verifying the stability of the learned controllers, are deferred to the supplementary materials due to space limit~\citep{lu2023bridging}.

\mysubsection{Validation of Robustness}
\label{sec:benchmark_robustness}

\begin{wrapfigure}{r}{0.5\textwidth}
  \small
  \setlength\tabcolsep{2pt} 
  \vspace{-2.5em}
  \captionsetup{type=table}
  \caption{\small Performance comparison on quadruple tank system with process noise and parametric uncertainties.}
  \vspace{-10pt}
  \label{tab:robustness}
  \begin{tabular}{c|ccccc}
    \toprule
    Method & Fail\% & Cost & P-Cost & Time(s) & \#Params \\
    \midrule
    MPC-T(16) & 82.6 & \textbf{216.8} & 713.4 & 0.25$_{\text{+0.56}}$ & -  \\
    Tube & 81.9 & \underline{233.3} & 597.9 & 2.22$_{\text{+44}}$ & -  \\
    Scenario & 16.4 & 236.9 & 273.2 & 5.21$_{\text{+18}}$ & -  \\
    RL-MLP & \textbf{1.3} & 238.9 & \textbf{241.5} & \underline{1$\times$10$^{-3}$} & 43K \\
    \textbf{LQP(4, 24)} & 1.5 & 256.7 & 261.8 & \textbf{2$\times$10$^{-4}$} & \textbf{0.3K}  \\
    \textbf{LQP(16, 96)} & \underline{1.4} & 240.6 & \underline{243.4} & 2$\times$10$^{-3}$ & \underline{2.7K} \\
    \bottomrule
  \end{tabular}
  \vspace{0.1em}

  {
    \footnotesize
   Notations are similar to those in the caption of Table~\ref{tab:common}. Computation time instead of FLOPs per control step is used as the metric for computational efficiency since it is difficult to obtain the exact FLOPs from the robust MPC baselines.
  }
   \vspace{10pt}
\end{wrapfigure}

This subsection is concerned with the robustness of the learned QP controller against modeling inaccuracies and disturbances. Instead of the nominal dynamics~\eqref{eq:nominal_sys}, we now consider the following perturbed dynamics:
\begin{equation*}
x_{k+1} = (A + \Delta A) x_k + (B + \Delta B) u_k + w_k,
\end{equation*}
where $\Delta A, \Delta B$ are parametric uncertainties, and $w_k$ is a disturbance. LQP and MLP are trained using domain randomization~\citep{tobin2017domain,mehta2020active}, where the simulator randomly sample these uncertain components during training. Robust MPC baselines, including tube MPC~\citep{mayne2005robust} and scenario MPC~\citep{bernardini2009scenario} implemented by the do-mpc toolbox~\citep{fiedler2023mpc}, are included for comparison.

The results in Table~\ref{tab:robustness} highlight LQP's robustness, as it achieves the success rate and constraint-violation-penalized cost comparable to MLP. Also, it requires significantly less online computation compared to robust MPC methods, benefitting from domain randomization known for its effectiveness in empirical RL and robotics~\citep{loquercio2019deep,margolis2022rapid}.

\mysubsection{Application Example on a Real-World System: Vehicle Drift Maneuvering}
\label{sec:benchmark_real}

\begin{figure}[!htbp]
  \centering
  \newlength\figureheight
  \newlength\figurewidth
  \setlength\figurewidth{0.45\textwidth}
  \setlength\figureheight{0.18\textwidth}
  \subfigure[\small Tracking performance]{
    \input{figures/qp_car_data.tex}
  }
  \subfigure[\small Blended frames of experiment video]{
      \includegraphics[width=0.45\textwidth]{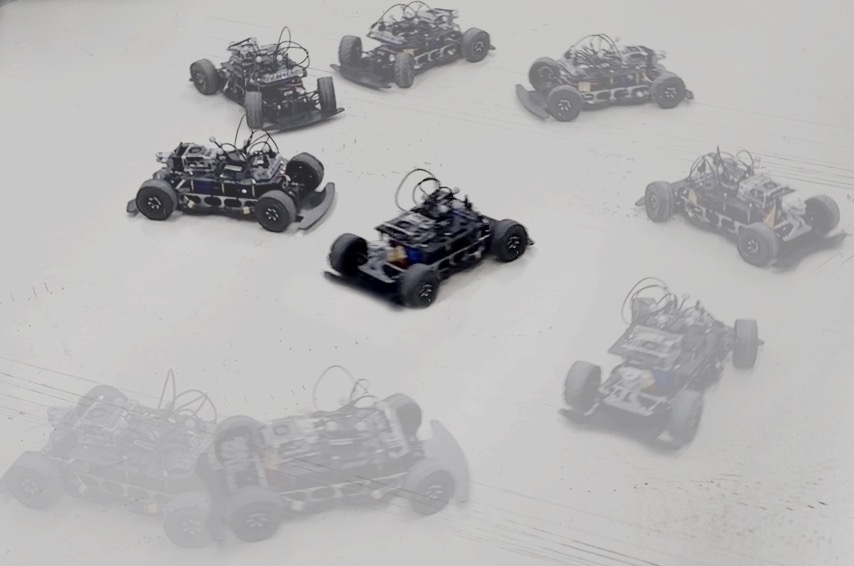}
  }
  \caption{\small Result of deploying learned QP controller to the vehicle drift maneuvering task. \revise{Video available at: \url{https://youtu.be/-XYtl2b4OVc}}.}
  \label{fig:car}
\end{figure}

LQP is also evaluated on a challenging robotics control task, namely, the drift maneuvering of a 1/10 scale RC car, similar to the problem studied in~\cite{yang2022hierarchical,domberg2022deep,lu2023consecutive}. The objective is to track the yaw rate, side slip angle, and velocity references, such that the car enters and maintains a drifting state. Despite the high nonlinearity of the system, the proposed controller formally introduced on linear systems successfully generalizes to this task. As shown in Fig.~\ref{fig:car}, the learned QP controller can track the references and maintain the drifting state, performing similarly to previous RL-trained MLP methods on this task~\citep{domberg2022deep}.\lina{did I mention that it would be good if we can add a link for the video}\yiwen{Sure! Link added.}

\vspace{-0.25em}

\mysection{Conclusion}
\label{sec:conclusion}

This work presents a novel class of QP controllers inspired by MPC. The proposed controllers not only retain the theoretical guarantees akin to MPC, but also exhibit desirable empirical performance and computational efficiency. Benchmarks including applications in real-world scenarios like vehicle drift maneuvering, further validate the effectiveness and robustness of our approach.

\revise{
Despite the promising results, several challenges remain: (i) the empirical results are not exhaustive nor conclusive, and further evidence is required to understand LQP's benefits and limitations in practice; (ii) the stability verification (Theorem~\ref{thm:stability}) still relies on the Lyapunov function of MPC, urging the simultaneous learning of policy and certificate similar to~\cite{chang2019neural}; (iii) the performance guarantees assume that the optimal QP solution is attained, a restriction that can potentially be lifted using techniques from~\cite{wu2022stability}.
}

Additionally, the unrolled QP solver is structured similarly to a deep neural network, indicating the suitability of the proposed policy architecture as a drop-in replacement for standard policy networks in RL. This opens up possibilities for combining the architecture with various RL methods, such as meta-learning~\citep{finn2017model} and safety-constrained RL~\citep{achiam2017constrained,yu2022reachability}, which are left for future investigation.\lina{it would be better to discuss a couple of specific limitations of the proposed approach. A habit of doing CS learning conference submissions, they always ask whether the paper discuss limitations.}\yiwen{Added discussion of the limitations.}


\newpage

\bibliography{refs}

\appendix

\newpage

\section{PDHG iterations and convergence guarantee}
\label{app:pdhg}

We introduce the PDHG iterations to solve Problem \ref{pb:QP} and analyze its iteration structure to facilitate further design of a QP-based controller. Define the following auxiliary objective functions
\begin{align}
f(z)=\mathbb{I}\{z \in \Rb^{m_{qp}}_{+}\},\ g(z)= \min_{y}\left\{ \frac{1}{2} y^{\top} P y+q^{\top} y | H y+b=z\right\} .
\end{align}
Then Problem \ref{pb:QP} is equivalent to the following problem where $z\in\Rb^{m_{qp}}$:
\begin{align*}
	\underset{z}{\minimize }\  f(z)+g(z) .
\end{align*}
The corresponding dual problem is
\begin{align*}
	\underset{\lambda}{\maximize}\  -f^*(-\lambda)-g^*(\lambda)
\end{align*}
with dual variable $\lambda\in\Rb^{m_{qp}}$, and $f^*,g^*$ are the Fenchel conjugate function of $f$ and $g$. The Lagrangian is
\begin{align}
	\bL(z,\lambda)=f(z)+\lambda^\top z - g^*(\lambda). \label{eq:Lag}
\end{align}
Since Problem \ref{pb:QP} has strongly convex objective due to $P\in\Sb^{n_{qp}}_{++}$ and affine constraints, strong duality holds.

Apply the variable metric proximal point method to the saddle subdifferential:
$$
\partial \bL(z, \lambda)=\left[\begin{array}{cc}
	0 & I^{\top} \\
	-I & 0
\end{array}\right]\left[\begin{array}{l}
	z \\
	\lambda
\end{array}\right]+\left[\begin{array}{c}
	\partial f(z) \\
	\partial g^*(\lambda)
\end{array}\right],\text{ and define }M=\left[\begin{array}{cc}
(1 / \alpha) I & I \\
I & (1 / \beta) I
\end{array}\right].
$$
Then the fix point iteration operator $(M+\partial {\bL})^{-1} M$ applied to iteration $k$ is
$$
\left[\begin{array}{cc}
	(1 / \alpha) I & 2I \\
	0 & (1/\beta)I
\end{array}\right]\left[\begin{array}{c}
	z^{k+1} \\
	\lambda^{k+1}
\end{array}\right]+\left[\begin{array}{c}
	\partial f\left(z^{k+1}\right) \\
	\partial g^*\left(\lambda^{k+1}\right)
\end{array}\right] \ni\left[\begin{array}{c}
	(1 / \alpha) z^k+  \lambda^k \\
	 z^k+(1/\beta) \lambda^k
\end{array}\right].
$$
The iteration can be equivalently written as:
\begin{align}
	& \lambda^{k+1}=\operatorname{Prox}_{ \beta g^*}\left(\lambda^k+\beta  z^k\right), \label{eq:PDHG_1} \\
	& z^{k+1}=\operatorname{Prox}_{\alpha f}\left(z^k-\alpha \left(2\lambda^{k+1}-\lambda^k \right) \right). \label{eq:PDHG_2}
\end{align}
The proximal operator of indicator function $f$ is expressed as $\operatorname{Prox}_{\alpha  f}(\cdot)=\proj_{\Rb^{m_{qp}}_+}(\cdot)$. We derive the explicit form of $\prox_{\beta g^*}(\cdot)$ in iteration \eqref{eq:PDHG_1}. According to definition,
\begin{align*}
	\prox_{ g/\beta }(\psi)=\underset{z}{\operatorname{argmin}}\left\{\min_{y}\frac{1}{2} y^{\top} P y+q^{\top} y+\frac{\beta}{2}\|z-\psi\|_2^2 \ \text{ s.t. } Hy+b=z\right\}.
\end{align*}
The KKT condition of the minimization problem associated with the proximal operator is
\begin{align}
	\left(\begin{array}{ccc}
		P & 0 & H^\top \\
		0 & \beta I & -I \\
		H & -I & 0
	\end{array}\right)\left(\begin{array}{l}
		y \\
		z \\
		d
	\end{array}\right)=\left(\begin{array}{c}
		-q \\
		\beta \psi \\
		-b
	\end{array}\right),\label{eq:KKT_prox}
\end{align}
where $d$ is the dual variable of equality constraint $Hy+b=z$.
Since $P\succ 0,\beta>0$, the square matrix on the LHS of \eqref{eq:KKT_prox} is non-singular and thus the solution $y$ to KKT condition \eqref{eq:KKT_prox} can be explicitly written as
\begin{align*}
	\prox_{ g/\beta }(\psi)=\beta H(\beta H^\top H+P)^{-1} H^\top \psi -F(HP^{-1} q -b),  \text{ with } F\triangleq (I+\beta HP^{-1}H^\top)^{-1}.
\end{align*}
As a result, by Moreau's Identity
$\phi=\operatorname{Prox}_{\beta g^*}(\phi)+\beta  \operatorname{Prox}_{g/\beta}( \phi/\beta),$
the PDHG iteration \eqref{eq:PDHG_1} can be written as
\begin{align}
	\lambda^{k+1}= F \lambda^k +\beta F z^k +\beta F(HP^{-1}q-b),\label{eq:PDHG_cal_1}
\end{align}
which is a simple affine transformation of $(z,\lambda)$.
Plugging $\lambda^{k+1}$ of \eqref{eq:PDHG_cal_1} into \eqref{eq:PDHG_2}, one obtains the following iteration
\begin{align}
	z^{k+1}&=\proj_{\Rb^{m_{qp}}_+} \left( (I-2\alpha\beta F)z^k+ \alpha(I-2F)\lambda^k -2\alpha\beta F (HP^{-1}q-b) \right) . \label{eq:PDHG_cal_2}
\end{align}

Due to the affine form of dual step \eqref{eq:PDHG_cal_1}, the $\lambda^{k+1}$ in \eqref{eq:PDHG_cal_2} can be expressed by affine transform of $z^k$ and $\lambda^k$. As a result, there is no $\lambda^{k+1}$ in update equation \eqref{eq:PDHG_cal_2}. Thus, the iterations \eqref{eq:PDHG_cal_1}-\eqref{eq:PDHG_cal_2} do not need to store history dual state $\lambda^k$ and current dual state $\lambda^{k+1}$ simultaneously for the calculation of $z^{k+1}$, which is an important advantage for efficient and memory-economic GPU tensor operations.

Since strong duality holds for Problem \ref{pb:QP} and thus Lagrange \eqref{eq:Lag}, we have the following convergence theorem for PDHG iterations. Define the primal and dual optimal solution to Lagrange \eqref{eq:Lag} as $z^\star,\lambda^\star$. The following result provides the convergence of PDHG \citep{ryu2022large}:
\begin{theorem}
	If $\alpha,\beta>0$, and $\alpha\beta<1$, then iterations \eqref{eq:PDHG_cal_1}-\eqref{eq:PDHG_cal_2} yield $z^k \rightarrow z^{\star}$ and $\lambda^k \rightarrow \lambda^{\star}$.
\end{theorem}

For simplicity of calculation and notations, we fix $\beta=1$ in our implementation \eqref{eq:pdhg_iter}.

\section{Ensuring the Feasibility of the Learned QP Problem}
\label{app:feasibility}

A caveat of the policy architecture described in Section~\ref{sec:method} is that the learned QP problem may not be feasible, since the sequence of PDHG iterations (i.e., recursively application of~\eqref{eq:pdhg_iter}) will return a vector $y^{{n}_{iter}}$, regardless of whether the QP problem parameterized by $(P, H, q, b)$ it attempts to solve is feasible. Even though the control input extracted from such $y^{{n}_{iter}}$ may lead to good empirical performance, the use of an infeasible QP as the control law is less desirable, because i) the explainability is compromised compared to deriving the control input from the optimal solution of a feasible QP; ii) as shown in Section~\ref{sec:theory}, a feasible QP can simplify the theoretical analysis of the controller.

To address this issue, we propose softening the learned constraints by introducing an additional slack variable $\epsilon$. Specifically, given learned parameters $(P, H, q, b)$, consider the following QP problem as a replacement for the original problem in~\eqref{eq:qp}:
  \begin{equation}
  \begin{aligned}
    \underset{y \in \mathbb{R}^{n_{qp}}, \epsilon \in \mathbb{R}}{\text{minimize}} &\quad \frac{1}{2} y^{\top} P y+q^{\top} y + \rho_\epsilon \epsilon^2, \\
    \text{subject to} &\quad H y+b+\epsilon \boldsymbol{1} \geq 0, \epsilon \geq 0,
  \end{aligned}
  \label{eq:slacked_qp}
  \end{equation}
  where $\rho_\epsilon > 0$ is a constant coefficient for penalizing a large slack variable.
Note that the new problem~\eqref{eq:slacked_qp} is always feasible regardless of the choice of $(P, H, q, b)$, since for any $(P, H, q, b)$ and any $y$, one can always choose a sufficiently large $\epsilon$ such that $(y, \epsilon)$ is a feasible solution to~\eqref{eq:slacked_qp}. With $\tilde{y} := [y^\top\; \epsilon^\top]^\top$ being the new decision variable, one can rearrange~\eqref{eq:slacked_qp} into the standard form~\eqref{eq:qp}, with the parameters being:
\begin{equation}
  \tilde{P} = \begin{bmatrix}
      P & 0 \\
      0 & 2\rho_\epsilon
      \end{bmatrix}, \quad
      \tilde{q} = \begin{bmatrix}
          q \\
          0
      \end{bmatrix}, \quad
      \tilde{H} = \begin{bmatrix}
          H & \mathbf{1} \\
          0 & 1
      \end{bmatrix}, \quad
      \tilde{b} = \begin{bmatrix}
          b \\
          0
      \end{bmatrix},
\end{equation}
Hereafter, with a slight abuse of notation, we always assume that the control policy solves a QP problem parameterized by $(\tilde{P}, \tilde{H}, \tilde{q}, \tilde{b})$ even when we denote the parameters as $(P, H, q, b)$.

\section{Intuition Behind Learned QP Controller: a Comparison with MPC}
\label{sec:example}

In this appendix, we exemplify why the learned controller can overcome the limitations of MPC via a toy example.

\newcommand{\citebbm}{\citep[p. 246, Example 12.1]{borrelli2017predictive}}
\begin{example}[Double integrator, variant of~\citebbm]
  Consider Problem~\ref{prob:infinite_horizon} with $n_{sys} = 2, m_{sys} = 1$, where
  \begin{equation}
    A = \begin{bmatrix}
      1 & 1 \\
      0 & 1
    \end{bmatrix},
    B = \begin{bmatrix}
      0 \\
      1
    \end{bmatrix},
    Q = I, R = 100,
    u_{\min} = -0.5, u_{\max} = 0.5,
    x_{\min} = \begin{bmatrix}
      -5 \\
      -5
    \end{bmatrix},
    x_{\max} = \begin{bmatrix}
      5 \\
      5
    \end{bmatrix}.
    \label{eq:double_integrator}
  \end{equation}
  The task is to stabilize the system around the origin, i.e., $r = 0$.
\end{example}

\newcommand{\citemci}{\citep[p. 192, Definition 10.10]{borrelli2017predictive}}
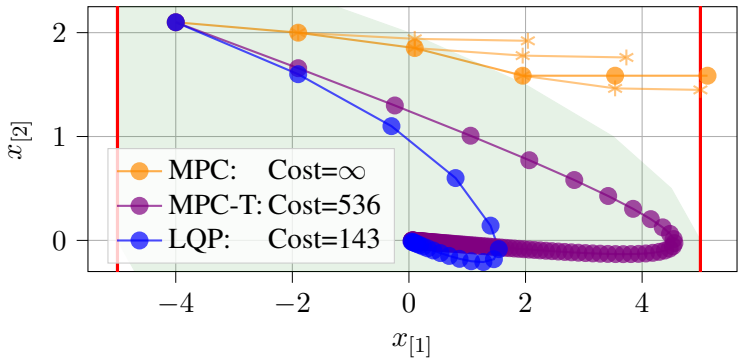
\begin{wrapfigure}{l}{0.67\textwidth}
  \setlength\figurewidth{0.67\textwidth}
  \setlength\figureheight{0.5\figurewidth}

  \centering
\begin{tikzpicture}

    \definecolor{darkgray176}{RGB}{176,176,176}
    \definecolor{darkorange}{RGB}{255,140,0}
    \definecolor{green01270}{RGB}{0,127,0}
    \definecolor{lightgray204}{RGB}{204,204,204}
    \definecolor{purple}{RGB}{128,0,128}

    \begin{axis}[
    height=\figureheight,
    legend cell align={left},
    legend style={
      fill opacity=0.8,
      draw opacity=1,
      text opacity=1,
      at={(0.03,0.03)},
      anchor=south west,
      draw=lightgray204
    },
    tick align=outside,
    tick pos=left,
    width=\figurewidth,
    x grid style={darkgray176},
    xlabel={\(\displaystyle x_{[1]}\)},
    xmajorgrids,
    xmin=-5.50610066927758, xmax=5.62811405482922,
    xtick style={color=black},
    y grid style={darkgray176},
    ylabel={\(\displaystyle x_{[2]}\)},
    ymajorgrids,
    ymin=-0.3, ymax=2.25,
    ytick style={color=black}
    ]
    \addplot [thick, darkorange, opacity=0.7, mark=*, mark size=3, mark options={solid}]
    table {%
    -4 2.1
    -1.9 1.9995860765537
    0.0995860765536969 1.85155801830395
    1.95114409485765 1.58543464534699
    3.53657874020464 1.58543464534699
    5.12201338555163 1.58543464534699
    };
    \addlegendentry{\makebox[3.2em][l]{MPC:} Cost=$\infty$}
    \path [draw=green01270, fill=green01270, opacity=0.1]
    (axis cs:-4,2.75)
    --(axis cs:-5,2.75)
    --(axis cs:-5,0)
    --(axis cs:-4.5,-0.5)
    --(axis cs:-3.5,-1)
    --(axis cs:-2,-1.5)
    --(axis cs:0,-2)
    --(axis cs:2.5,-2.5)
    --(axis cs:4,-2.75)
    --(axis cs:5,-2.75)
    --(axis cs:5,0)
    --(axis cs:4.5,0.5)
    --(axis cs:3.5,1)
    --(axis cs:2,1.5)
    --(axis cs:0,2)
    --(axis cs:-2.5,2.5)
    --cycle;
    \addplot [thick, purple, opacity=0.7, mark=*, mark size=3, mark options={solid}]
    table {%
    -4 2.1
    -1.9 1.65789663087797
    -0.242103369122028 1.29885564843611
    1.05675227931408 1.00764978317287
    2.06440206248695 0.77182724490132
    2.83622930738827 0.58120653130726
    3.41743583869553 0.427463160885437
    3.84489899958097 0.303791604980963
    4.14869060456193 0.204628736790942
    4.35331934135287 0.125427604607321
    4.4787469459602 0.0624723728104315
    4.54121931877063 0.0127269402729286
    4.55394625904356 -0.0262888914739237
    4.52765736756963 -0.056600713040577
    4.47105665452905 -0.0798623656484493
    4.39119428888061 -0.0974237178740528
    4.29377057100655 -0.110386104923025
    4.18338446608353 -0.119647675386171
    4.06373679069736 -0.125940481970576
    3.93779630872678 -0.129860818555473
    3.80793549017131 -0.131894032573819
    3.67604145759749 -0.132434818105443
    3.54360663949205 -0.131803812137545
    3.4118028273545 -0.130261166801099
    3.2815416605534 -0.128017647971668
    3.15352401258173 -0.125243710476729
    3.02828030210501 -0.122076918226154
    2.90620338387885 -0.118628010563007
    2.78757537331584 -0.11498586130618
    2.67258951200966 -0.111221532106102
    2.56136797990356 -0.107391585044832
    2.45397639485873 -0.103540789397647
    2.35043560546108 -0.0997043329200353
    2.25073127254105 -0.0959096279383987
    2.15482164460265 -0.0921777860917155
    2.06264385851093 -0.0885248221301433
    1.97411903638079 -0.0849626361809911
    1.8891564001998 -0.0814998148977148
    1.80765658530208 -0.0781422845495435
    1.72951430075254 -0.0748938430902331
    1.65462045766231 -0.0717565933206557
    1.58286386434165 -0.0687312952321674
    1.51413256910948 -0.0658176523229363
    1.44831491678655 -0.0630145439842684
    1.38530037280228 -0.0603202138493571
    1.32498015895292 -0.0577324221935313
    1.26724773675939 -0.0552485689999884
    1.2119991677594 -0.0528657930984304
    1.15913337466097 -0.0505810517971204
    1.10855232286385 -0.0483911846216655
    1.06016113824219 -0.0462929641136156
    1.01386817412857 -0.0442831361019909
    0.96958503802658 -0.0423584514192393
    0.92722658660734 -0.0405156906719702
    0.88671089593537 -0.0387516833814728
    0.847959212553897 -0.0370633225675262
    0.810895889986371 -0.0354475756515446
    0.775448314334827 -0.0339014923936582
    0.741546821941168 -0.0324222104463471
    0.709124611494821 -0.031006958999369
    0.678117652495452 -0.029653060902544
    0.648464591592908 -0.0283579335809135
    0.620106658011995 -0.0271190889979209
    0.592987569014074 -0.025934132874178
    0.567053436139896 -0.0248007633301139
    0.542252672809782 -0.0237167690887447
    0.518535903721037 -0.0226800273486429
    0.495855876372395 -0.0216885014158401
    0.474167374956554 -0.0207402381659988
    0.453427136790556 -0.0198333653940055
    0.43359377139655 -0.0189660890965918
    0.414627682299958 -0.0181366907241927
    0.396490991575766 -0.0173435244306117
    0.379147467145154 -0.016585014342861
    0.362562452802293 -0.0158596518685132
    0.34670280093378 -0.0151659930538273
    0.331536807879952 -0.0145026560026222
    0.31703415187733 -0.0138683183632129
    0.303165833514117 -0.0132617148885934
    0.289904118625524 -0.0126816350733361
    0.277222483552188 -0.012126920869312
    0.265095562682876 -0.0115964644812523
    0.253499098201624 -0.0110892062423116
    0.242409891959312 -0.0106041325691238
    0.231805759390188 -0.0101402739953172
    0.221665485394871 -0.00969670328205892
    0.211968782112812 -0.00927253360389495
    0.202696248508917 -0.0088669168079304
    0.193829331700987 -0.00847904174423771
    0.185350289956749 -0.00810813266527257
    0.177242157291476 -0.00775344769201176
    0.169488709599465 -0.00741427734449398
    0.162074432254971 -0.00708994313443661
    0.154984489120534 -0.00677979621761443
    0.14820469290292 -0.0064832161037141
    0.141721476799205 -0.00619960942141869
    0.135521867377787 -0.00592840873652598
    0.129593458641261 -0.00566907142096023
    0.123924387220301 -0.00542107857059802
    0.118503308649703 -0.00518393396989324
    0.113319374679809 -0.00495716310135217
    0.108362211578457 -0.00474031219797737
    0.10362189938048 -0.00453294733686635
    0.0990889520436134 -0.00433465357221877
    0.0947542984713947 -0.00414503410607207
    0.0906092643653226 -0.00396370949515141
    0.0866455548701712 -0.00379031689228334
    0.0828552379778879 -0.00362450932088547
    0.0792307286570024 -0.00346595498110496
    0.0757647736758974 -0.00331433658623737
    0.07245043708966 -0.00316935072811469
    0.0692810863615454 -0.00303070727020578
    0.0662503790913396 -0.0028981287672261
    0.0633522503241135 -0.00277134991010408
    0.0605809004140094 -0.00265011699520093
    0.0579307834188085 -0.00253418741672766
    0.0553965960020808 -0.00242332918134866
    0.0529732668207322 -0.00231732044400437
    0.0506559463767278 -0.00221594906402754
    0.0484399973127003 -0.00211901218066747
    };
    \addlegendentry{\makebox[3.2em][l]{MPC-T:} Cost=536}
    \addplot [thick, blue, opacity=0.7, mark=*, mark size=3, mark options={solid}]
    table {%
    -4 2.1
    -1.9 1.6
    -0.3 1.1
    0.8 0.6
    1.4 0.142012542486191
    1.54201254248619 -0.0842916905879973
    1.45772085189819 -0.18107175976038
    1.27664909213781 -0.208336427435279
    1.06831266470254 -0.200695571117103
    0.867617093585432 -0.177605147100985
    0.690011946484447 -0.149596015550196
    0.540415930934251 -0.122032674588263
    0.418383256345988 -0.09736057985574
    0.321022676490248 -0.0764325750991701
    0.244590101391078 -0.059279747866094
    0.185310353524984 -0.0455488363280891
    0.139761517196895 -0.0347427014261483
    0.105018815770747 -0.0263458505272864
    0.0786729652434603 -0.0198845040053128
    0.0587884612381475 -0.0149503649212419
    0.0438380963169056 -0.0112052649259566
    };
    \addlegendentry{\makebox[3.2em][l]{LQP:} Cost=143}
    \addplot [thick, darkorange, opacity=0.5, mark=asterisk, mark size=3, mark options={solid}, forget plot]
    table {%
    -1.9 1.9995860765537
    0.0995860765536969 1.9405653892967
    2.0401514658504 1.92135187059097
    };
    \addplot [thick, darkorange, opacity=0.5, mark=asterisk, mark size=3, mark options={solid}, forget plot]
    table {%
    0.0995860765536969 1.85155801830395
    1.95114409485765 1.77885699205218
    3.73000108690983 1.76124454658796
    };
    \addplot [thick, darkorange, opacity=0.5, mark=asterisk, mark size=3, mark options={solid}, forget plot]
    table {%
    1.95114409485765 1.58543464534699
    3.53657874020464 1.46342125979534
    4.99999999999998 1.44893194039142
    };
    \addplot [very thick, red, forget plot]
    table {%
    -5 -5
    5 -5
    };
    \addplot [very thick, red, forget plot]
    table {%
    5 -5
    5 5
    };
    \addplot [very thick, red, forget plot]
    table {%
    5 5
    -5 5
    };
    \addplot [very thick, red, forget plot]
    table {%
    -5 5
    -5 -5
    };
    \end{axis}

\end{tikzpicture}
  \caption{\small Comparison of trajectories under different controllers on the double integrator example, starting from the common initial state $x_0 = (-4, 2.1)$. ``MPC'' stands for the truncated MPC, ``MPC-T'' stands for MPC with a manually crafted terminal cost, and ``LQP'' stands for the learned QP controller. Solid dots represent realized trajectories, while asterisks stand for predicted trajectories. The green shadow stands for the maximal control invariant set~\citemci, i.e., the largest set over which one can expect \emph{any} controller to work. The thick red lines stand for the bounds on the state.
  }
  \label{fig:toy_example}
\end{wrapfigure}

We consider MPC with horizon $N=3$. We also train a QP controller that solves a problem of the same dimension, i.e., $n_{qp} = Nm_{sys} = 3, m_{qp} = N(m_{sys} + n_{sys}) = 9$, for sake of fair comparison.

A notable feature of this example is the high cost of control, from which one can expect the MPC to exert a weak control effort, as long as the state does not go out of bounds within its prediction horizon. This is the case with the orange trajectory in Fig.~\ref{fig:toy_example}: its ``flatness'' indicates a small local cost if we focus our attention on a $N$-long segment of the trajectory excluding the last step (since variations along the $x_{[2]}$-axis would require costly control inputs), but the trajectory as a whole is unacceptable because it violates the state constraints. The myopia of optimizing the cost over such a short horizon can be mitigated by introducing a terminal cost, as shown by the purple trajectory in Fig.~\ref{fig:toy_example}, which incorporates a terminal cost $50 x_{[2]}^2$. However, manually crafting of a terminal cost is nontrivial, and it is unclear how to optimize this design against the true objective, i.e., the cumulative cost along the entire trajectory before the state is brought to the origin. By comparison, the learned QP controller is free of myopia or manual crafting, since the RL algorithm naturally considers entire trajectories, and finds a QP-based control policy within the class described in Section~\ref{sec:method} that can generate the trajectories with lowest expected total cost.

\section{Proofs of Performance Guarantees and Numerical Examples}
\label{app:theory}

\subsection{Proof of Theorem~\ref{thm:feasibility}}

\begin{proof}
  According to Appendix~\ref{app:feasibility}, the optimization problem in~\eqref{eq:qp_opt} is always feasible. Therefore, we only need to prove that the control input $\pi_\theta(x)$ provided by~\eqref{eq:qp_opt} always keeps the system state in a feasible region, a sufficient condition for which is:
  \begin{equation}
    G(Ax + B\pi_\theta(x)) \leq c \text{  whenever  } Gx \leq c.
    \label{eq:invariance}
  \end{equation}
  The condition~\eqref{eq:invariance} is equivalent to the following condition:
  \begin{equation}
    \max_{x: Gx \leq c} \min_{\epsilon \in \mathbb{R}} \left\{ \epsilon \mid G (Ax + B \pi_\theta(x)) - c \leq \epsilon \right\} \leq 0.
    \label{eq:maxmin}
  \end{equation}
  According to the strong duality of linear programming,
  \begin{equation}
    \min_{\epsilon \in \mathbb{R}} \left\{ \epsilon \mid G (Ax + B \pi_\theta(x)) - c \leq \epsilon \right\} = \max_{\nu \in \mathbb{R}^p} \left\{ \nu^\top(G (Ax + B \pi_\theta(x)) - c) \mid \nu \geq 0, \mathbf{1}^\top \nu = 1 \right\},
    \label{eq:strong_duality}
  \end{equation}
  where $p$ is the dimension of $c$.
  Substituting~\eqref{eq:strong_duality} into~\eqref{eq:maxmin}, and flipping maximization to minimization, we obtain that the condition~\eqref{eq:invariance} is equivalent to the optimal solution of the following problem being nonnegative:
  \begin{subequations}
  \begin{align}
    \underset{x, \nu}{\text{minimize}} &\quad -\nu^\top(G (Ax + B \pi_\theta(x)) - c), \\
    \text{subject to} &\quad \nu \geq 0, \mathbf{1}^\top \nu = 1,
  \end{align}
  \end{subequations}
  Expanding $\pi_\theta(x)$ using~\eqref{eq:qp_opt}, we obtain the following equivalent problem:
  \begin{subequations}
  \begin{align}
    \underset{x, \nu}{\text{minimize}} &\quad -\nu^\top(G (Ax + B y_{[1:m_{sys}]} ) - c), \\
    \text{subject to} &\quad \nu \geq 0, \mathbf{1}^\top \nu = 1, \\
    &\quad y^* \in \operatorname{arg\,min} \left\{ (1/2) y^\top P y + q^\top y \mid Hy + b \geq 0 \right\}.
    \label{eq:inner_problem_in_feasibility}
  \end{align}
  \end{subequations}
  According to Appendix~\ref{app:feasibility}, the inner problem~\eqref{eq:inner_problem_in_feasibility} is feasible and satisfies Slater's condition. Therefore,~\eqref{eq:inner_problem_in_feasibility} can be equivalently replaced by its KKT condition:
  \begin{equation}
    Py + W_q x_0  - H^\top \mu = 0,
    Hy + W_b x_0 + b_b \geq 0, \mu \geq 0,
    \mu^\top(Hy + W_b x_0 + b_b) = 0,
    \label{eq:kkt_in_feasibility}
  \end{equation}
  where $\mu$ is the dual variable of proper dimensions associated with the inequality constraint in~\eqref{eq:inner_problem_in_feasibility}. Substituting~\eqref{eq:kkt_in_feasibility} into~\eqref{eq:inner_problem_in_feasibility}, we obtain the problem in the theorem statement, whose optimal solution being nonnegative is sufficient for persistent feasibility.
\end{proof}

\subsection{Proof of Theorem~\ref{thm:stability}}

Here we formulate the baseline MPC required for Theorem~\ref{thm:stability}:

\begin{problem}[Stabilizing Baseline MPC with horizon $N$]\label{pb:baseline_MPC}
    \begin{subequations}
    	\begin{align}
    		\underset{x_{1:N}, u_{0:N-1}}{\text{minimize}} &\quad J(x_0, u_{0:N-1}) = \sum_{k=0}^{N-1} l(x_k, u_k) + V_f(x_N), \\
    		 \text{subject to} \quad & \quad x_{k+1} = A x_{k} + B u_k, \quad k = 0, \cdots, {N-1},  \\
        & \quad x_{N} \in \mathcal{X}_f,
    	\end{align}
    \end{subequations}
    where:
    \begin{itemize}
      \item $l(x, u) = x^\top Q x + u^\top R u$ is the stage cost;
      \item $V_f(x) = x^\top P_f x$ is the quadratic cost-to-go function under a stabilizing linear feedback controller $u = Kx$;
      \item $\mathcal{X}_f$ is a positive invariant set under the same linear feedback controller $u = Kx$.
    \end{itemize}
    Here $K$ can be any stabilizing linear feedback gain, e.g., it can be chosen as the LQR gain determined by $(A, B, Q, R)$.
\end{problem}

Theorem~\ref{thm:stability} can be proved by comparing the Lyapunov decrease of the learned controller and the above defined stabilizing baseline MPC:

\begin{proof}
  Consider the following candidate Lyapunov function:
  \begin{equation}
    V(x_0) = \min_{\bar{u}_{0:N}} \left\{ J(x_0, \bar{u}_{0:N}) \mid x_{N+1}(Ax_0+Bu_0, \bar{u}_{1:N}) \in \mathcal{X}_f \right\},
    \label{eq:lyap_candidate}
  \end{equation}
  where
  \begin{equation}
    J(x_0, \bar{u}_{0:N}) = \sum_{i=0}^{N} l(x_i, \bar{u}_i) + V_f(x_{N+1})\text{ when $x_{0:N+1}$ is the state trajectory under $\bar{u}_{0:N}$},
    \label{eq:J_def}
  \end{equation}
  and $u_0 = y_{[1:m_{sys}]}$ stands for the control input given by the learned policy~\eqref{eq:qp_opt} when the state is $x_0$,
  and $x_{N+1}(Ax_0+Bu_0, \bar{u}_{1:N})$ stands for the state at time step $N+1$ if the control input sequence $\bar{u}_{1:N}$ is applied to the system starting from $x_1 = Ax_0 + Bu_0$. Since the baseline MPC is well-defined, $V(x)$ takes finite values over $\mathcal{X}_0$. One can verify that i) $u_0 = \min\{\frac12 y^\top P y | Hy + b_b \geq 0\}_{[1:m_{sys}]} = 0$ when $x_0 = 0$ and hence $V(0) = J(0, 0) = 0$; ii) $V(x_0) \geq l(x_0, \bar{u}_0^*) \geq x_0^\top Q x_0 > 0$ for all $x_0 \neq 0$, where $\bar{u}_0^*$ is first control input from the optimal solution to the problem in~\eqref{eq:lyap_candidate}. Therefore, we only need to verify that $V(Ax_0 + Bu_0) - V(x_0) < 0$ for all $x_0 \in \mathcal{X}_0 \backslash \{0\}$.

  Consider the following problem:
  \begin{subequations}
  \begin{align}
  \underset{x_0, \bar{u}_{0:N}, u_0, y}{\text{minimize}} & \quad J(x_0, \bar{u}_{0:N}) + l(x_0, u_0) - J(x_0, (u_0, \bar{u}_{1:N})) - \epsilon \| x_0 \|^2, \label{eq:stability_prob_obj} \\
  \text{subject to} & \quad Gx_0 \leq c,
   x_{N+1}(Ax_0+Bu_0, \bar{u}_{1:N}) \in \mathcal{X}_f, \\
   & u_0 = y_{[1:m_{sys}]}, \quad y \in \operatorname{arg\,min} \left\{ (1/2) y^\top P y + q^\top y \mid Hy + b \geq 0 \right\}, \label{eq:stability_prob_inner}
  \end{align}
  \end{subequations}
  According to Appendix~\ref{app:feasibility}, the inner problem~\eqref{eq:stability_prob_inner} is feasible and satisfies Slater's condition. Therefore,~\eqref{eq:stability_prob_inner} can be equivalently replaced by its KKT condition, and the bilevel problem~\eqref{eq:stability_prob_obj}-\eqref{eq:stability_prob_inner} is equivalent to the one in the theorem statement. Since the optimal value of this problem is nonnegative, it holds for any $x_0 \in \mathcal{X}_0$ that:
  \begin{align}
    0 &\leq J(x_0, \bar{u}_{0:N}^*(x_0)) + l(x_0, u_0) - J(x_0, (u_0, \bar{u}_{1:N}^*(x_0))) - \epsilon \| x_0 \|^2 \nonumber \\
    &= V(x_0) + l(x_0, u_0) - J(x_0, (u_0, \bar{u}_{1:N}^*(x_0))) - \epsilon \| x_0 \|^2,
    \label{eq:ineq_VJ}
  \end{align}
  where $\bar{u}^*(x_0)$ stands for the state sequence that achieves the optimal value of the problem in~\eqref{eq:lyap_candidate} when the initial state is $x_0$.

  Now assume that the system state starts from an arbitrary $x_0 \in \mathcal{X}_0$, the control input $u_0$ given by the learned controller is applied for the first step. Further assume that the control input sequence $\bar{u}^*_{1:N}$ is applied for the subsequent $N$ steps, i.e., $u_{1:N} = \bar{u}^*_{1:N}$, and that at step $N+1$, the control input $u_{N+1} = Kx_{N+1}$ is applied, where $K$ is the stabilizing linear feedback gain defined in Problem~\ref{pb:baseline_MPC}~\footnote{Note that $u_{1:N+1}$ are only defined here for deriving the Lyapunov decrease, but they will not be actually applied to the system, since $u_0$ given by the learned problem is updated in a receding horizon manner at every step.}. Since $x_{N+1} \in \mathcal{X}_f$ by definition of $\bar{u}^*_{1:N}$, and $\mathcal{X}_f$ is invariant under the controller $u = Kx$, we have $x_{N+2}(x_0, u_{0:N+1}) \in \mathcal{X}_f$. Hence,
  \begin{flalign*}
    V(x_1) &\leq \sum_{i=1}^{N+1} l(x_i, u_i) + V_f(x_{N+2}) & \text{(Optimality in definition of $V$)}\\
    & = \sum_{i=0}^{N} l(x_i, u_i) - l(x_0, u_0) + l(x_{N+1}, u_{N+1}) + V_f(x_{N+2}) & \text{(Shifting summation index)}\\
    & = \sum_{i=0}^{N} l(x_i, u_i) - l(x_0, u_0) + V_f(x_{N+1}) & \text{(Definition of $V_f$)} \\
    &= J(x_0, u_{0:N}) - l(x_0, u_0) &  \text{(Definition of $J$~\eqref{eq:J_def})} \\
    & \leq V(x_0) - \epsilon \| x_0 \|^2, & \text{(Inequality~\eqref{eq:ineq_VJ})}.
  \end{flalign*}
Hence, we have $V(x_1) - V(x_0) < 0$ for every $x_0 \in \mathcal{X}_0\backslash \{ 0 \}$, which verifies the stability of the closed-loop system under the learned controller.
\end{proof}

\subsection{A Numerical Example}

Consider the double integrator example in Appendix~\ref{sec:example}. Due to the innate symmetry of the system, we consider a symmetrized parameterization of the learned QP controller~\eqref{eq:qp_opt}:
\begin{align}
  & \pi_\theta(x_0) = y^*_{[1:m_{sys}]}, y^* \in \operatorname{arg\,min} \left\{ (1/2) y^\top P y + q^\top y \mid -1 \leq \hat{H} y \leq 1 \right\}, \nonumber\\
  & \text{where } q = W_q x_0.
  \label{eq:qp_sym}
\end{align}
Furthermore, the technique introduced in Appendix~\ref{app:feasibility} is applied to $(P, q, H)$ with penalty coefficient $\rho_\epsilon = 10$ to ensure the feasibility of the learned QP problem. The overall problem size is chosen to be $n_{qp} = 4, m_{qp} = 20$ (such that~\eqref{eq:qp_sym} has 10 pairs of symmetrized inequality constraints).
A set of parameters resulting from training the controller using the PPO algorithm~\citep{schulman2017proximal} is:

\begin{footnotesize}
\begin{align}
  & P = \begin{bmatrix}
    \num{71.693}    & \num{-59.678}  & \num{-6.376}   &  0.0 \\
    \num{-59.678}   & \num{50.029}   & \num{5.334}    &  0.0 \\
    \num{-6.376}    & \num{5.334}    & \num{0.695}    &  0.0 \\
    0.0             & 0.0            & 0.0            & 10.0 \\
  \end{bmatrix},
  W_q = \begin{bmatrix}
    \num{0.416}     & \num{0.159} \\
    \num{-0.265}    & \num{0.408} \\
    \num{-0.020}    & \num{0.023} \\
    0.0             & 0.0 \\
  \end{bmatrix}, \label{eq:PWq} \\
  & \hat{H} = \begin{bmatrix}
    \num{0.183} & \num{0.568} & \num{-0.109} & \num{-0.839} & \num{0.377} & \num{0.285} & \num{0.174} & \num{0.356} & \num{0.632} & 0.0 \\
    \num{0.034} & \num{-0.116} & \num{0.149} & \num{0.371} & \num{-0.858} & \num{-0.265} & \num{-0.427} & \num{-0.408} & \num{-0.657} & 0.0 \\
    \num{-1.518} & \num{-0.686} & \num{0.925} & \num{-1.335} & \num{-0.652} & \num{-0.643} & \num{-1.839} & \num{-1.337} & \num{-1.630} & 0.0 \\
    1.0 & 1.0 & 1.0 & 1.0 & 1.0 & 1.0 & 1.0 & 1.0 & 1.0 & 1.0 \\
  \end{bmatrix}^\top.
\end{align}
\end{footnotesize}

Casting the learned parameters back to the original parameterization~\eqref{eq:qp_opt}, the controller parameters $P, W_q$ are the same as~\eqref{eq:PWq}, while $H, W_b, b_b$ can be obtained as follows:
\begin{equation}
  H = \begin{bmatrix}
    \hat{H}^\top & -\hat{H}^\top
  \end{bmatrix}^\top, W_b = 0, b_b = \mathbf{1}_{m_{qp}}.
  \label{eq:HWb}
\end{equation}

We first verify the persistent feasibility of the learned controller on some initial set. To guess a suitable initial set, we first estimate the maximal control invariant set of the system by recursively computing the backward reachable set starting from the original. The resulting estimate of maximal control invariant set for the system~\eqref{eq:double_integrator} is $\{x | \hat{G} x \leq \hat{c} \}$, where:

\begin{footnotesize}
\begin{equation}
  \begin{bmatrix}\hat{G}^\top \\ \hat{c}^\top\end{bmatrix} = \left[\begin{array}{@{}c@{\hspace{2mm}}c@{\hspace{2mm}}c@{\hspace{2mm}}c@{\hspace{2mm}}c@{\hspace{2mm}}c@{\hspace{2mm}}c@{\hspace{2mm}}c@{\hspace{2mm}}c@{\hspace{2mm}}c@{\hspace{2mm}}c@{\hspace{2mm}}c@{\hspace{2mm}}c@{\hspace{2mm}}c@{\hspace{2mm}}c@{\hspace{2mm}}c@{\hspace{0mm}}}
    \num{0} & \num{1} & \num{0.71} & \num{-1} & \num{-0.71} & \num{0} & \num{0.45} & \num{-0.24} & \num{-0.45} & \num{-0.32} & \num{0.2} & \num{0.16} & \num{0.24} & \num{0.32} & \num{-0.16} & \num{-0.2} \\
    \num{-1} & \num{0} & \num{0.71} & \num{0} & \num{-0.71} & \num{1} & \num{0.89} & \num{-0.97} & \num{-0.89} & \num{-0.95} & \num{0.98} & \num{0.99} & \num{0.97} & \num{0.95} & \num{-0.99} & \num{-0.98} \\
    \num{2.8} & \num{5} & \num{3.5} & \num{5} & \num{3.5} & \num{2.8} & \num{2.5} & \num{2.0} & \num{2.5} & \num{2.1} &  \num{2.0} & \num{2.1} & \num{1.9} & \num{2.1} & \num{2.1} & \num{2.0}
  \end{array}\right].
\end{equation}
\end{footnotesize}

A guess of the initial set can be obtained by shrinking the estimated maximal control invariant set slightly. In particular, we guess the initial set $\{ x | Gx \leq c \}$, where:
\begin{equation}
  G = \hat{G}, c = \hat{c} - 0.2,
  \label{eq:initial_set}
\end{equation}
where ``$-$'' stands for element-wise subtraction.

With parameters defined in~\eqref{eq:PWq}, \eqref{eq:HWb} and \eqref{eq:initial_set}, the feasibility verification problem in Theorem~\ref{thm:feasibility} is fully instantiated.
In what follows, we find the upper and lower bounds to this problem, with all computation done on a workstation with 16-Core Intel i7-13700KF CPU.
Solving the problem using the local nonlinear programming solver IPOPT~\citep{biegler2009large}, which finishes in 0.4 seconds, we quickly obtain an \emph{upper bound} to the optimal value: $p^* \leq 0.0828$. Note that this is only a hint that the controller may be persistently feasible on the initial set $\{ x | Gx \leq c \}$, but not a certificate, since the upper bound provided by a local solver may not be tight. To obtain a certificate, we use the \texttt{SumOfSquares.jl} toolbox in Julia~\citep{legat2017sos,weisser2019polynomial} to relax this nonconvex QCQP problem to a SDP problem, and solve the SDP problem using the Mosek solver~\citep{aps2023mosek}. The solver finishes in 216 seconds, returning a \emph{lower bound} to the optimal value of the original problem: $p^* \geq 0.0803$. This nonnegative lower bound certifies that the learned controller is indeed persistently feasible on the initial set $\{ x | Gx \leq c \}$, and the lower and upper bounds are quite close in this example.

We next verify the asymptotic stability of the learned controller on the same initial set $\{ x | Gx \leq c \}$. For the current problem, it suffices to choose baseline MPC horizon $N = 0$, which reduces the verification problem in Theorem~\ref{thm:stability} to finding a quadratic Lyapunov function $V_f(x) = x^\top P_f x$ such that the following problem has nonnegative optimal value:
\begin{subequations}
\begin{align}
  \underset{x_0, y, \mu}{\text{minimize}} &\quad V_f(x_0) - V_f(Ax_0 + By_{[1:m_{sys}]} ) - \epsilon \| x_0 \|^2, \label{eq:simplified_stability_1} \\
  \text{subject to} &\quad Gx_0 \leq c, \\
  & Py + W_q x_0  - H^\top \mu = 0, Hy + W_b x_0 + b_b \geq 0, \mu \geq 0, \mu^\top(Hy + W_b x_0 + b_b) = 0. \label{eq:simplified_stability_3}
\end{align}
\end{subequations}
To find such a Lyapunov function, we first approximate the control policy $\pi(x)$ defined in~\eqref{eq:qp_sym} by a linear feedback controller $u = \hat{K} x$ near the origin: by choosing $x_1 = \begin{bmatrix}
  0.1 & 0
\end{bmatrix}^\top$ and $x_2 = \begin{bmatrix}
  0 & 0.1
\end{bmatrix}^\top$, we obtain:
\begin{equation}
  \hat{K} = \begin{bmatrix}
    \pi(x_1) & \pi(x_2)
  \end{bmatrix} \begin{bmatrix}
    x_1 & x_2
  \end{bmatrix}^{-1} = \begin{bmatrix}
    -0.20  -1.28
  \end{bmatrix},
\end{equation}
and setting $P_f$ to be the solution to the Lyapunov equation $(A+B\hat{K}) P_f (A+B\hat{K})^\top + Q + \hat{K}^\top R \hat{K} = 0$, we obtain:
\begin{equation}
  P_f = \begin{bmatrix}
    5.64 &  12.59 \\
    12.59 &  58.40
  \end{bmatrix}.
  \label{eq:Pf}
\end{equation}

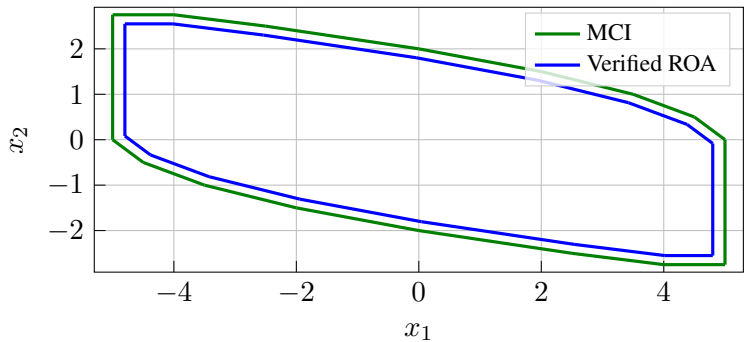
\begin{wrapfigure}{l}{0.67\textwidth}

  \setlength\figurewidth{0.67\textwidth}
  \setlength\figureheight{0.5\figurewidth}
\begin{tikzpicture}

    \definecolor{green}{RGB}{0,128,0}
    \definecolor{lightgray204}{RGB}{204,204,204}
    
    \begin{axis}[
    height=\figureheight,
    legend cell align={left},
    legend style={fill opacity=0.8, draw opacity=1, text opacity=1, draw=lightgray204},
    tick pos=left,
    width=\figurewidth,
    xlabel=\textcolor{black}{\(\displaystyle x_1\)},
    xmajorgrids,
    xmin=-5.3, xmax=5.3,
    xtick style={color=black},
    xtick={-4,-2,0,2,4},
    xticklabels={
      \(\displaystyle -4\),
      \(\displaystyle -2\),
      \(\displaystyle 0\),
      \(\displaystyle 2\),
      \(\displaystyle 4\)
    },
    ylabel=\textcolor{black}{\(\displaystyle x_2\)},
    ymajorgrids,
    ymin=-2.915, ymax=2.915,
    ytick style={color=black},
    ytick={-2,-1,0,1,2},
    yticklabels={
      \(\displaystyle -2\),
      \(\displaystyle -1\),
      \(\displaystyle 0\),
      \(\displaystyle 1\),
      \(\displaystyle 2\)
    }
    ]
    \addplot [very thick, green]
    table {%
    -4.5 -0.5
    -3.5 -1
    };
    \addlegendentry{\footnotesize MCI}
    \addplot [very thick, green, forget plot]
    table {%
    -3.5 -1
    -1.99999999999998 -1.50000000000001
    };
    \addplot [very thick, green, forget plot]
    table {%
    -1.99999999999998 -1.50000000000001
    0 -2
    };
    \addplot [very thick, green, forget plot]
    table {%
    0 -2
    2.49999999999997 -2.5
    };
    \addplot [very thick, green, forget plot]
    table {%
    2.49999999999997 -2.5
    4 -2.75
    };
    \addplot [very thick, green, forget plot]
    table {%
    4 -2.75
    5 -2.75
    };
    \addplot [very thick, green, forget plot]
    table {%
    5 -2.75
    5 6.14892752100087e-16
    };
    \addplot [very thick, green, forget plot]
    table {%
    5 6.14892752100087e-16
    4.5 0.5
    };
    \addplot [very thick, green, forget plot]
    table {%
    4.5 0.5
    3.5 1
    };
    \addplot [very thick, green, forget plot]
    table {%
    3.5 1
    2 1.5
    };
    \addplot [very thick, green, forget plot]
    table {%
    2 1.5
    0 2
    };
    \addplot [very thick, green, forget plot]
    table {%
    0 2
    -2.50000000000001 2.5
    };
    \addplot [very thick, green, forget plot]
    table {%
    -2.50000000000001 2.5
    -4 2.75
    };
    \addplot [very thick, green, forget plot]
    table {%
    -4 2.75
    -5 2.75
    };
    \addplot [very thick, green, forget plot]
    table {%
    -5 2.75
    -5 5.46571335200076e-16
    };
    \addplot [very thick, green, forget plot]
    table {%
    -5 5.46571335200076e-16
    -4.5 -0.5
    };
    \addplot [very thick, blue]
    table {%
    -4.38152817055072 -0.335629116974661
    -3.42327027756748 -0.814758063466282
    };
    \addlegendentry{\footnotesize{Verified ROA}}
    \addplot [very thick, blue, forget plot]
    table {%
    -3.42327027756748 -0.814758063466282
    -1.94404124723588 -1.30783440691015
    };
    \addplot [very thick, blue, forget plot]
    table {%
    -1.94404124723588 -1.30783440691015
    0.0438900147434397 -1.80481722240498
    };
    \addplot [very thick, blue, forget plot]
    table {%
    0.0438900147434397 -1.80481722240498
    2.53606088601309 -2.30325139665891
    };
    \addplot [very thick, blue, forget plot]
    table {%
    2.53606088601309 -2.30325139665891
    4.01655250605964 -2.55
    };
    \addplot [very thick, blue, forget plot]
    table {%
    4.01655250605964 -2.55
    4.8 -2.55
    };
    \addplot [very thick, blue, forget plot]
    table {%
    4.8 -2.55
    4.8 -0.0828427124746193
    };
    \addplot [very thick, blue, forget plot]
    table {%
    4.8 -0.0828427124746193
    4.38152817055072 0.33562911697466
    };
    \addplot [very thick, blue, forget plot]
    table {%
    4.38152817055072 0.33562911697466
    3.42327027756747 0.814758063466283
    };
    \addplot [very thick, blue, forget plot]
    table {%
    3.42327027756747 0.814758063466283
    1.94404124723589 1.30783440691014
    };
    \addplot [very thick, blue, forget plot]
    table {%
    1.94404124723589 1.30783440691014
    -0.0438900147434298 1.80481722240497
    };
    \addplot [very thick, blue, forget plot]
    table {%
    -0.0438900147434298 1.80481722240497
    -2.53606088601314 2.30325139665892
    };
    \addplot [very thick, blue, forget plot]
    table {%
    -2.53606088601314 2.30325139665892
    -4.01655250605964 2.55
    };
    \addplot [very thick, blue, forget plot]
    table {%
    -4.01655250605964 2.55
    -4.8 2.55
    };
    \addplot [very thick, blue, forget plot]
    table {%
    -4.8 2.55
    -4.8 0.0828427124746195
    };
    \addplot [very thick, blue, forget plot]
    table {%
    -4.8 0.0828427124746195
    -4.38152817055072 -0.335629116974661
    };
    \end{axis}
    
\end{tikzpicture}
    
  \caption{\small Comparison of boundaries of the Maximal Control Invariant (MCI) set and verified Region Of Attraction (ROA) for the double integrator example.}
  \label{fig:verified_roa}
\end{wrapfigure}

Plugging~\eqref{eq:Pf} into the problem~\eqref{eq:simplified_stability_1}-\eqref{eq:simplified_stability_3}, we obtain a fully defined stability verification problem. Similarly to the feasibility verification problem, we try to solve the stability verification using both the local solve IPOPT and the global solver based on \texttt{SumOfSquares.jl} and Mosek. The local solver finishes in 0.007 seconds, which the global solver finishes in 1.4 seconds, both returning the optimal value $p^* = 0$ up to numerical accurary $10^{-5}$ (the optimal value is attained at $x_0 = 0$). This verifies the asymptotic stability of the learned controller on the initial set $\{ x | Gx \leq c \}$.

The maximal control invariant set $\{ x | \hat{G}x \leq \hat{c} \}$ and verified region of attraction $\{ x | Gx \leq c \}$ for this example are visualized in Fig.~\ref{fig:verified_roa}. The verified region of attraction is only slightly smaller than the maximal control invariant set (accounts for 92\% of the area of the maximal control invariant set), which demonstrates the effectiveness of our proposed verification method.

\section{Benchmarking Setup and Additional Experiment Results}
\label{app:benchmark}

\subsection{Environment Definition for RL Training}
\label{app:rl}

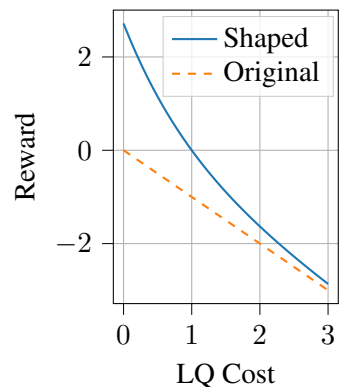
\begin{wrapfigure}{r}{0.3\textwidth}

  \setlength\figurewidth{0.3\textwidth}
  \setlength\figureheight{1.2\figurewidth}
\begin{tikzpicture}

\definecolor{darkgray176}{RGB}{176,176,176}
\definecolor{darkorange25512714}{RGB}{255,127,14}
\definecolor{lightgray204}{RGB}{204,204,204}
\definecolor{steelblue31119180}{RGB}{31,119,180}

\begin{axis}[
height=\figureheight,
legend cell align={left},
legend style={fill opacity=0.8, draw opacity=1, text opacity=1, draw=lightgray204},
tick align=outside,
tick pos=left,
width=\figurewidth,
x grid style={darkgray176},
xlabel={LQ Cost},
xmajorgrids,
xmin=-0.15, xmax=3.15,
xtick style={color=black},
y grid style={darkgray176},
ylabel={Reward},
ymajorgrids,
ymin=-3.28591409142295, ymax=3.004195919882,
ytick style={color=black}
]
\addplot [thick, steelblue31119180]
table {%
0 2.71828182845905
0.0303030303030303 2.60684217304774
0.0606060606060606 2.49782432389104
0.0909090909090909 2.39115599371392
0.121212121212121 2.2867670529078
0.151515151515152 2.18458946512746
0.181818181818182 2.08455722481028
0.212121212121212 1.98660629656046
0.242424242424242 1.89067455634242
0.272727272727273 1.79670173442968
0.303030303030303 1.70462936005644
0.333333333333333 1.61440070772134
0.363636363636364 1.52596074509394
0.393939393939394 1.43925608247612
0.424242424242424 1.35423492377195
0.454545454545455 1.27084701892108
0.484848484848485 1.18904361775192
0.515151515151515 1.10877742521218
0.545454545454545 1.03000255793577
0.575757575757576 0.952674502106079
0.606060606060606 0.876750072576983
0.636363636363636 0.802187373214042
0.666666666666667 0.728945758419423
0.696969696969697 0.656985795805273
0.727272727272727 0.586269229981222
0.757575757575758 0.516758947422776
0.787878787878788 0.448418942388341
0.818181818181818 0.381214283853568
0.848484848484849 0.315111083432658
0.878787878787879 0.250076464257172
0.909090909090909 0.186078530783755
0.939393939393939 0.123086339503062
0.96969696969697 0.0610698705229713
1 0
1.03030303030303 -0.0601515266064001
1.06060606060606 -0.119412120481228
1.09090909090909 -0.177808374626828
1.12121212121212 -0.235366088284415
1.15151515151515 -0.292110290626642
1.18181818181818 -0.348065263743001
1.21212121212121 -0.403254564939122
1.24242424242424 -0.457701048370487
1.27272727272727 -0.511426886030399
1.3030303030303 -0.564453588111505
1.33333333333333 -0.616802022759544
1.36363636363636 -0.668492435237485
1.39393939393939 -0.719544466517639
1.42424242424242 -0.769977171318829
1.45454545454545 -0.819809035605173
1.48484848484848 -0.869057993562556
1.51515151515152 -0.917741444068385
1.54545454545455 -0.965876266669736
1.57575757575758 -1.01347883708459
1.60606060606061 -1.06056504224036
1.63636363636364 -1.10715029486359
1.66666666666667 -1.15324954763407
1.6969696969697 -1.19887730691664
1.72727272727273 -1.2440476460829
1.75757575757576 -1.28877421843552
1.78787878787879 -1.33307026974659
1.81818181818182 -1.37694865042183
1.84848484848485 -1.42042182730167
1.87878787878788 -1.4635018951101
1.90909090909091 -1.50620058756178
1.93939393939394 -1.54852928813763
1.96969696969697 -1.5904990405388
2 -1.63212055882856
2.03030303030303 -1.67340423727148
2.06060606060606 -1.71436015987899
2.09090909090909 -1.75499810966993
2.12121212121212 -1.79532757765474
2.15151515151515 -1.83535777155147
2.18181818181818 -1.87509762424162
2.21212121212121 -1.91455580197365
2.24242424242424 -1.95374071232148
2.27272727272727 -1.99266051190562
2.3030303030303 -2.03132311388374
2.33333333333333 -2.06973619521761
2.36363636363636 -2.10790720372326
2.39393939393939 -2.14584336491062
2.42424242424242 -2.183551688619
2.45454545454545 -2.22103897545454
2.48484848484848 -2.25831182303553
2.51515151515152 -2.29537663205126
2.54545454545455 -2.33223961214008
2.57575757575758 -2.36890678759197
2.60606060606061 -2.40538400288091
2.63636363636364 -2.44167692803213
2.66666666666667 -2.47779106382911
2.6969696969697 -2.51373174686523
2.72727272727273 -2.54950415444459
2.75757575757576 -2.58511330933654
2.78787878787879 -2.6205640843883
2.81818181818182 -2.65586120699997
2.84848484848485 -2.69100926346582
2.87878787878788 -2.72601270318616
2.90909090909091 -2.76087584275339
2.93939393939394 -2.7956028699161
2.96969696969697 -2.8301978474249
3 -2.86466471676339
};
\addlegendentry{Shaped}
\addplot [thick, darkorange25512714, dashed]
table {%
0 -0
0.0303030303030303 -0.0303030303030303
0.0606060606060606 -0.0606060606060606
0.0909090909090909 -0.0909090909090909
0.121212121212121 -0.121212121212121
0.151515151515152 -0.151515151515152
0.181818181818182 -0.181818181818182
0.212121212121212 -0.212121212121212
0.242424242424242 -0.242424242424242
0.272727272727273 -0.272727272727273
0.303030303030303 -0.303030303030303
0.333333333333333 -0.333333333333333
0.363636363636364 -0.363636363636364
0.393939393939394 -0.393939393939394
0.424242424242424 -0.424242424242424
0.454545454545455 -0.454545454545455
0.484848484848485 -0.484848484848485
0.515151515151515 -0.515151515151515
0.545454545454545 -0.545454545454545
0.575757575757576 -0.575757575757576
0.606060606060606 -0.606060606060606
0.636363636363636 -0.636363636363636
0.666666666666667 -0.666666666666667
0.696969696969697 -0.696969696969697
0.727272727272727 -0.727272727272727
0.757575757575758 -0.757575757575758
0.787878787878788 -0.787878787878788
0.818181818181818 -0.818181818181818
0.848484848484849 -0.848484848484849
0.878787878787879 -0.878787878787879
0.909090909090909 -0.909090909090909
0.939393939393939 -0.939393939393939
0.96969696969697 -0.96969696969697
1 -1
1.03030303030303 -1.03030303030303
1.06060606060606 -1.06060606060606
1.09090909090909 -1.09090909090909
1.12121212121212 -1.12121212121212
1.15151515151515 -1.15151515151515
1.18181818181818 -1.18181818181818
1.21212121212121 -1.21212121212121
1.24242424242424 -1.24242424242424
1.27272727272727 -1.27272727272727
1.3030303030303 -1.3030303030303
1.33333333333333 -1.33333333333333
1.36363636363636 -1.36363636363636
1.39393939393939 -1.39393939393939
1.42424242424242 -1.42424242424242
1.45454545454545 -1.45454545454545
1.48484848484848 -1.48484848484848
1.51515151515152 -1.51515151515152
1.54545454545455 -1.54545454545455
1.57575757575758 -1.57575757575758
1.60606060606061 -1.60606060606061
1.63636363636364 -1.63636363636364
1.66666666666667 -1.66666666666667
1.6969696969697 -1.6969696969697
1.72727272727273 -1.72727272727273
1.75757575757576 -1.75757575757576
1.78787878787879 -1.78787878787879
1.81818181818182 -1.81818181818182
1.84848484848485 -1.84848484848485
1.87878787878788 -1.87878787878788
1.90909090909091 -1.90909090909091
1.93939393939394 -1.93939393939394
1.96969696969697 -1.96969696969697
2 -2
2.03030303030303 -2.03030303030303
2.06060606060606 -2.06060606060606
2.09090909090909 -2.09090909090909
2.12121212121212 -2.12121212121212
2.15151515151515 -2.15151515151515
2.18181818181818 -2.18181818181818
2.21212121212121 -2.21212121212121
2.24242424242424 -2.24242424242424
2.27272727272727 -2.27272727272727
2.3030303030303 -2.3030303030303
2.33333333333333 -2.33333333333333
2.36363636363636 -2.36363636363636
2.39393939393939 -2.39393939393939
2.42424242424242 -2.42424242424242
2.45454545454545 -2.45454545454545
2.48484848484848 -2.48484848484848
2.51515151515152 -2.51515151515152
2.54545454545455 -2.54545454545455
2.57575757575758 -2.57575757575758
2.60606060606061 -2.60606060606061
2.63636363636364 -2.63636363636364
2.66666666666667 -2.66666666666667
2.6969696969697 -2.6969696969697
2.72727272727273 -2.72727272727273
2.75757575757576 -2.75757575757576
2.78787878787879 -2.78787878787879
2.81818181818182 -2.81818181818182
2.84848484848485 -2.84848484848485
2.87878787878788 -2.87878787878788
2.90909090909091 -2.90909090909091
2.93939393939394 -2.93939393939394
2.96969696969697 -2.96969696969697
3 -3
};
\addlegendentry{Original}
\end{axis}

\end{tikzpicture}
  \caption{\small Visualization of the reward shaping scheme used (with hyperparameters $\rho_{sta} = 1, c_1 = 1, c_2 = 1$): a positive reward is added to small LQ cost, in order to encourage the controller to minimize the steady-state cost.}
  \label{fig:reward_shaping}
\end{wrapfigure}

All learned QP controllers are trained using Reinforcement Learning (RL). To simulate the constrained optimal control problem described in Problem~\ref{prob:infinite_horizon}, the RL environment consists of components described as follows:

\begin{itemize}
  \item \textbf{Observation}: the observation variable is $x = [x_0^\top, r^\top]^\top$ when training policies for tracking a reference $r$, and $x = x_0$ when training policies for stabilizing the system around the origin.
  \item \textbf{Action Preprocessing}: given policy output $u_0$, the actual action applied to the system is chosen to be $u = \operatorname{clip}(u_0, u_{\min}, u_{\max})$, where $\operatorname{clip}(x, a, b) = \min(\max(x, a), b)$ element-wise. In other words, the action is clipped to the bounds on the control input, even if the policy output is outside of the bounds.
  \item \textbf{Reward}: let $(x, u, x')$ be a state, action, and next state triple. Recall from~\eqref{eq:cost} that the LQ cost is defined as $l(x', u) = (x' - r)^\top Q (x' - r) + u^\top R u$. Taking into account the original LQ cost, the state constraint, and a \mbox{reward shaping term for minimizing the steady-state cost, the} overall reward function for RL training is chosen as:
    \begin{align}
    \mathrm{Reward}(x, u, x') = & - l(x', u) && \text{(Original LQ cost)} \nonumber\\
    & - \rho_{pen} \left( 1 - \mathbf{1}_{ \left\{ x_{\min} \leq x'  \leq x_{\max} \right\}} \right) && \text{(Penalty on constraint violation)} \nonumber\\
    & - \rho_{sta} \exp(-c_1(l(x', u) - c_2 )) && \text{(Reward shaping term)},
    \label{eq:reward_shaping}
    \end{align}
    where $\rho_{sta}, c_1 > 0, c_2$ are hyperparameters for reward shaping. See Fig.~\ref{fig:reward_shaping} for a visualization of the reward shaping scheme.

  \end{itemize}

  \begin{itemize}
  \item \textbf{Episode Definition}: at the start of an episode, both the initial state and the reference are randomly sampled according to a task-specific distribution (see ``Initial State'' and ``Reference'' in Table~\ref{tab:tasks}), and an episode is terminated on either of the following conditions:
  \begin{itemize}
    \item The state constraint is violated;
    \item The episode length exceeds a predefined threshold (see ``Episode length'' in Table~\ref{tab:tasks}).
  \end{itemize}

  \item \textbf{Domain randomization}: when training the controllers for the nominal systems (subsection~\ref{sec:benchmark_nominal}), no domain randomization is used, i.e., the environment simulates the deterministic dynamics~\eqref{eq:nominal_sys}. On the other hand, when training the robust controllers for uncertain systems (subsection~\ref{sec:benchmark_robustness}), the environment simulates the following perturbed dynamics:
  \begin{equation}
    x_{k+1} = (A + \Delta A) x_k + (B + \Delta B) u_k + w_k,
  \end{equation}
  where two sources of randomness are incorporated into the dynamics:
  \begin{itemize}
    \item Perturbation in the system parameters: $\Delta A, \Delta B$ are randomly sampled from a task-specific distribution at the start of each episode (see ``System Perturbation'' in Table~\ref{tab:tasks});
    \item Process noise: $w_k$ is sampled from a task-dependent distribution at each time step (see ``Process Noise'' in Table~\ref{tab:tasks}).
  \end{itemize}
\end{itemize}

\subsection{Task Definitions}

The details on the benchmarking tasks are summarized in Table~\ref{tab:tasks}.

\newcommand{\narrowminus}{\resizebox{0.5em}{\height}{$-$}}

  {\small
  \centering
  \begin{longtable}{|p{0.07\textwidth}|p{0.19\textwidth}|p{0.19\textwidth}|p{0.19\textwidth}|p{0.19\textwidth}|}
    \caption{\small Task definitions for benchmarking.}
    \label{tab:tasks}
    \\
    \hline
    & \textbf{Double Integrator} & \textbf{Quadruple Tank} & \textbf{Cartpole Balance} & \textbf{Drift Control} \\
    \hline
    \endfirsthead

    \hline
    & \textbf{Double Integrator} & \textbf{Quadruple Tank} & \textbf{Cartpole Balance} & \textbf{Drift Control} \\
    \hline
    \endhead

    \textbf{Source} & \citebbm & \cite{johansson2000quadruple} (original source); \cite{li2023linear} (linearization) & \cite{geva1993cartpole} &  \cite{lu2023consecutive} \\
    \hline
    \textbf{Remark} & Classical example where improperly designed MPC may fail. & Stable linearized system with adjustable reference. & Unstable system with slight nonlinearity. & Highly nonlinear robotics control task. \\
    \hline
    \textbf{Nominal dynamics} & $
    \begin{aligned}
    & A = \begin{bmatrix}
      1 & 1 \\
      0 & 1
    \end{bmatrix}, \\
    & B = \begin{bmatrix}
      0 \\
      1
    \end{bmatrix}.
    \end{aligned}
    $
    &
    {\scriptsize
    $
    \begin{aligned}
      & A = \left[\begin{array}{@{}c@{}c@{}c@{}c@{}}
        0.98 & 0 & 0.04 & 0 \\
        0 & 0.99 & 0 & 0.03 \\
        0 & 0 & 0.96 & 0 \\
        0 & 0 & 0 & 0.97
      \end{array}\right] \\
      & B = \left[\begin{array}{@{}c@{}c@{}}
        0.83 & 0 \\
        0 & 0.62 \\
        0 & 0.47 \\
        0.3 & 0
      \end{array}\right].
    \end{aligned}
    $
    }
    &
    {\scriptsize
    \vspace{-5em}
    $
    \left[\begin{array}{@{}c@{\hspace{1mm}}c@{}}
      m_c + m_p & m_p l c_\theta \\
      m_p l c_\theta & m_p l^2
    \end{array}\right]
    \begin{bmatrix}
      \ddot{p}_x \\ \ddot{\theta}
    \end{bmatrix} \allowbreak = \begin{bmatrix}
      u + m_p l s_\theta \dot{\theta}^2 \\
      m_p g l s_\theta \\
    \end{bmatrix},
    $
    where state variable is $x = [p_x, \dot{p}_x, \theta, \dot{\theta}]^\top$, and nominal parameters are $m_c = 1, m_p = 0.1, l = 0.55$.

    Discretized with time step 0.1s.

    Simulation uses nonlinear dynamics, while baseline MPC uses linearized dynamics.
    }
    &
    {\scriptsize
    \vspace{-5em}

    \cite[Eq. (5)]{lu2023consecutive}, with intermediate variables determined by \cite[Eq. (6)]{lu2021two}.

    State variable is $x = [r, \beta, V]^\top$, input variable is $u = [\delta, \omega]^\top$ and nominal parameters are $B=0.9, C=2.13, D=0.382$.

    Discretized with time step 0.1s.

    Nonlinear dynamics is used.
    } \\
    \hline
    \textbf{Con-straints} & $-5 \leq x \leq 5$, $-0.5 \leq u \leq 0.5$. & $0 \leq x \leq 20$,\qquad $0 \leq u \leq 8$. & $-2 \leq p_x \leq 2$, $-0.5 \leq \theta \leq 0.5$, $-10 \leq u \leq 10$. & $-7.6 \leq r \leq 7.6$, $-3.8 \leq \beta \leq 3.8$, $0 \leq V \leq 4$, \qquad $-0.6 \leq \delta \leq 0.6$, $1\leq \omega \leq 7$. \\
    \hline
    \textbf{Cost matrices} & $Q = I, R = 100I$. & $Q = I, R = 0.1I$. & $Q = \operatorname{diag}([1, \allowbreak 10^{-4}, \allowbreak 1, \allowbreak 10^{-4}]), \allowbreak R = 10^{-4}$. & $Q = I$,  $R = \operatorname{diag}([1, 0.01])$. \\
    \hline
    \textbf{Initial state} & Uniformly sampled over state space, then projected to maximal control invariant set. & Uniform~over~$[0, 16]^4$ (ensures that initial state is within maximal control invariant set). & $p_x \sim \mathcal{U}[-1.8, 1.8], \allowbreak \dot{p}_x \sim \mathcal{U}[-1, 1],\allowbreak \theta \sim \mathcal{U}[-0.1, 0.1],\allowbreak \dot{\theta} \sim \mathcal{U}[-0.1, 0.1]$. & $r=\beta=V=0$ (starts from still). \\
    \hline
    \textbf{Refe-rence} & $x = 0$ (stabilizing around the origin). & Uniform over state space. & $p_x \sim \mathcal{U} [-2, 2]$, $\dot{p}_x = \theta = \dot{\theta} = 0$ (balancing pole at random position). & $r \sim s \times \mathcal{U}[1.5, 2.5]$, $\beta \! \sim \!-s \!\times \!\mathcal{U}[0.8, 1.2]$, $V \sim \mathcal{U}[1.5, 2.5]$, $s\sim \mathcal{U}\{ -1, 1 \}$ (Circular drifting path in either CW or CCW direction). \\
    \hline
    \textbf{Episode length} & 100 & 500 & 100 & 1500 \\
    \hline
    \textbf{System perturbation} & $\Delta A = \begin{bmatrix}
      \delta_1 & \delta_2 \\
      0 & \delta_3
    \end{bmatrix}$, $\Delta B = \begin{bmatrix}
      0 \\ \delta_4
    \end{bmatrix}$,
    $\delta_i \stackrel{i.i.d.}{\sim} \mathcal{U}[-0.05, 0.05]$ &
    {
    \scriptsize
    $\Delta A =  \left[\begin{array}{@{}c@{}c@{}c@{}c@{}}
      \delta_1 & 0 & \delta_3 & 0 \\
      0 & \delta_2 & 0 & \delta_4 \\
      0 & 0 & \narrowminus\delta_1 & 0 \\
      0 & 0 & 0 & \narrowminus\delta_2
    \end{array}\right]$, $\Delta B = \left[\begin{array}{@{}c@{}c@{}}
      8.3\delta_5 & 0 \\
      0 & 6.2\delta_6 \\
      0 & 6.2\delta_6 \\
      3\delta_5 & 0
    \end{array}\right]$,
    $\delta_i \! \stackrel{i.i.d.}{\sim} \!0.002\!\times\! \mathcal{U}[-1, 1]$.
    } & $m_c \sim \mathcal{U}[0.7, 1.3]$, $m_p \sim \mathcal{U}[0.07, 0.13]$, $l \sim \mathcal{U}[0.4, 0.7]$.
    & $B \sim \mathcal{U}[0.8, 1]$, $C \sim \mathcal{U}[2, 2.5]$, $D \sim \mathcal{U}[0.3, 0.4]$. \\
    \hline
    \textbf{Process noise} & $w_k \! \stackrel{i.i.d.}{\sim} \!\mathcal{N}(0, 0.05I)$. & $w_k \! \stackrel{i.i.d.}{\sim} \!\mathcal{N}(0, 0.1I)$. & $w_k \stackrel{i.i.d.}{\sim} \mathcal{N}(0, \allowbreak \operatorname{diag}([0, 0.1, 0, 0.1]))$ & {
      \scriptsize
      Disturbance on $f_{ij}$ defined in \cite[Eq. (5)]{lu2023consecutive}, sampled from AR(1) process with parameter 0.97 and noise standard deviation 0.03.
    }
    \\
    \hline
  \end{longtable}
  }

\subsection{Training Details}

All the training (MLP and LQP) is done using the Proximal Policy Optimization (PPO) algorithm~\citep{schulman2017proximal}.
The same set of hyperparameters is used for all the experiments, except for the size of the function approximators (number of neurons for MLP, and number of variables/constraints for LQP), since multiple sizes are tested for each task.
The hyperparameters used for training are summarized in Table~\ref{tab:ppo_hyperparams}.

\begin{table}[!htbp]
  \caption{\small Hyperparameters used for training.}
  \label{tab:ppo_hyperparams}
  \centering
\begin{tabular}{@{}ll@{}}
  \toprule
  \textbf{Parameter} & \textbf{Value} \\
  \midrule
  \textit{Common} & \\
  \quad Number of epochs & 5000 \\
  \quad Optimizer & Adam~\citep{kingma2014adam} \\
  \quad Critic learning rate & Linear annealing 1e-3 $\to$ 2e-6 \\
  \quad Actor learning rate & Linear annealing 5e-4 $\to$ 1e-6 \\
  \quad Entropy coefficient & 0 \\
  \quad Reward discount factor ($\gamma$) & 0.99 \\
  \quad Target smoothing parameter ($\tau$) & 0.95 \\
  \quad Batch size & 100000 \\
  \quad Horizon length & 20 \\
  \quad PPO mini-epochs & 1 \\
  \quad PPO clip & 0.2 \\
  \quad Penalty on constraint violation ($\rho_{pen}$, see~\eqref{eq:reward_shaping}) & 100000 \\
  \quad Reward shaping (see~\eqref{eq:reward_shaping}) & $\rho_{sta} =50, c_1 = 0.05, c_2 = 2$ \\
  \midrule
  \textit{MLP} & \\
  \quad Number of hidden layers & 3 \\
  \quad Number of neurons in hidden layers & $[4n, 2n, n]$, where $n \in \{ 8, 16, 32, 64 \}$ \\
  \quad Activation function & ELU \\
  \midrule
  \textit{LQP} & \\
  \quad Number of variables/constraints $(n_{qp}, m_{qp})$ & $(n, 6n)$, where $n \in \{ 4, 8, 16 \}$ \\
  \quad Number of unrolled iterations ($n_{iter}$) & 10 \\
  \quad PDHG step size ($\alpha$) & 1 \\
  \quad Residual regularization coefficient ($\rho_{res}$) & 1e-3 \\
  \quad Penalty for infeasibility ($\rho_\epsilon$ in Appendix~\ref{app:feasibility}) & 10 \\
  \bottomrule
\end{tabular}
\end{table}

\subsection{Definition of MPC Baselines}

In this subsection, we provide the details on the MPC baselines used in the benchmarking experiments. For each MPC baseline appearing in Tables~\ref{tab:common} and~\ref{tab:robustness}, the problem formulation (either in explicit formulas or provided by an existing software package) and the solver used for solving the MPC problem are summarized in Table~\ref{tab:mpc_baselines}.

\begin{table}[!htbp]
  \caption{\small Definition of MPC baselines.}
  \label{tab:mpc_baselines}
  \centering
  \begin{tabular}{@{}lll@{}}
    \toprule
    \textbf{Name} & \textbf{Formulation} & \textbf{Solver} \\
    \midrule
    MPC($N$) & Problem~\ref{pb:MPC}, translated to QP via~\eqref{eq:mpc2qp}-\eqref{eq:AB}. & OSQP~\citep{stellato2020osqp} \\[2ex]
    MPC-T($N$) & \begin{minipage}{8cm}Problem~\ref{pb:MPC} with terminal cost $\rho\| x_N \|^2$ added to the objective, translated to QP similarly to~\eqref{eq:mpc2qp}-\eqref{eq:AB}. Weight $\rho$ grid-searched over $\{ 1, 10, 100\}$, picking the best-performing one.\end{minipage} & OSQP~\citep{stellato2020osqp}  \\[6ex]
    Tube & \begin{minipage}{8cm} Tube MPC implemented by open-sourced code of \cite{doff2022difference}, simplified for linear systems. Horizon $N=16$ and terminal weight $\rho = 10$. Tube size grid-searched over $\{ 0.05, 0.1, 0.2, 0.3 \}$, since using the actual upper bound on disturbance magnitude may result in infeasibility.\end{minipage} & Mosek~\citep{aps2022mosek} \\[10ex]
    Scenario & \begin{minipage}{8cm}Scenario-based MPC implemented by the do-mpc toolbox~\citep{fiedler2023mpc}. Horizon $N = 16$ and terminal weight $\rho = 10$. Three scenarios are generated for each of $A, B, w$. \end{minipage} & IPOPT~\citep{biegler2009large} \\
    \bottomrule
  \end{tabular}
\end{table}

\subsection{Evaluation Metrics}
\label{app:metric}

In this subsection, we provide the details on the evaluation metrics used in the benchmarking experiments. For each controller on each task, we run $N_t = 10000$ independent trials/episodes. For each trial $i = 1, \ldots, N_t$, we denote the episode length as $T^{(i)}$, which can either be the maximum episode length listed in Table~\ref{tab:tasks} or smaller (due to early termination caused by constraint violation). We denote the state and input trajectories of the $i$-th trial as $\{ x_k^{(i)} \}_{k=1}^{T^{(i)}}$ and $\{ u_k^{(i)} \}_{k=1}^{T^{(i)}}$, respectively. The random generators for initial state, reference, system perturbation and process noise are seeded separately, such that for different controllers on the same task, the $i$-th trial always has the same initial state, reference, system perturbation and realization of process noise. The metrics appearing in Tables~\ref{tab:common} and~\ref{tab:robustness} are defined as follows:
\begin{itemize}
  \item \textbf{Fail\%}: percentage of trials that are early-terminated due to constraint violation, i.e.,
  \begin{equation}
    \text{Fail\%} = \frac{1}{N_t} \sum_{i=1}^{N_t} \mathbf{1}_{\left\{ T^{(i)} < T_{\max} \right\}} \times 100,
  \end{equation}
  where $T_{\max}$ is the task-dependent maximum episode length listed in Table~\ref{tab:tasks}.
  \item \textbf{Cost}: average LQ cost per time step, i.e.,
  \begin{equation}
    \text{Cost} = \frac{1}{\sum_{i = 1}^{N_t}T^{(i)} } \sum_{i=1}^{N_t}  \sum_{k=1}^{T^{(i)}} l(x_k^{(i)}, u_k^{(i)}),
  \end{equation}
  where $l(x, u)$ is the LQ cost defined in~\eqref{eq:cost}.
  \item \textbf{P-Cost}: average LQ cost per time step, including the penalty for constraint violation, i.e.,
  \begin{equation}
    \text{P-Cost} = \frac{1}{\sum_{i = 1}^{N_t}T^{(i)} } \sum_{i=1}^{N_t}  \sum_{k=1}^{T^{(i)}} l(x_k^{(i)}, u_k^{(i)}) + \rho_{pen} \left( 1 - \mathbf{1}_{ \left\{ x_{\min} \leq  x_k^{(i)} \leq x_{\max} \right\}} \right),
  \end{equation}
  where $\rho_{pen}$ is the penalty for constraint violation defined in Table~\ref{tab:ppo_hyperparams}.
  \item \textbf{FLOPs}: number of Floating Point Operations (FLOPs) required for each time step $k$. For MLP and LQP, this value is constant since the path of forward propagations is fixed. For MPC, this value is variable since the number of iterations required for solving the MPC problem to the desired accuracy at each step is variable, and is reported in the format $\text{median}_{+(\text{max} - \text{median})}$.
  \item \textbf{Time}: computation time per step. For MLP and LQP, the time is measured by the time of batch operation on an NVIDIA RTX 4090 GPU. For MPC and robust MPC methods, the time is measured by solving a single instance of the MPC problem on a single core of an Intel i7-13700KF CPU. Overhead may apply since the testing program is parallelized. The time is reported in the format $\text{median}_{+(\text{max} - \text{median})}$ (subscript omitted when fluctuations are negligible).
  \item \textbf{\#Params}: number of learnable policy parameters, i.e., the total size of weight matrices and bias vectors in the MLP policy, or the total size of $W_q, W_b, b_b$ defined in~\eqref{eq:pdhg_affine_param} in the LQP policy. Note that critic parameters are excluded since the comparison is mainly concerned with the \emph{deployment} of learned policies.
\end{itemize}

\subsection{Complete Benchmarking Results}

In this subsection, we present extended versions of Tables~\ref{tab:common} and~\ref{tab:robustness}, which include additional configurations of the methods, as well as additional tasks.

The benchmarking results for nominal systems are presented in Table~\ref{tab:extended1}, which extends Table~\ref{tab:common} by including results for different weights of terminal cost in MPC, for MLPs with different sizes, and for an additional size of LQP. It also includes complete benchmarking results for the double integrator example. One can observe from the table that the proposed LQP method consistently achieves control performance similar to MPC with sufficiently long horizon and proper terminal cost, and MLP with sufficiently large size, while being more efficient in terms of FLOPs and number of learnable policy parameters.

\begin{table}[!htbp]
  \caption{\small Performance comparison on nominal systems (extended version).}\label{tab:extended1}
  \small
  \vspace{-10pt}\setlength\tabcolsep{3pt} 
  \renewcommand{\arraystretch}{1.0} 
  \begin{tabular}{>{\centering\arraybackslash}m{2.4cm}|ccccc|ccccc}
    \toprule
    \multirow{2}{*}{\diagbox[dir=NW,innerwidth=2.4cm,height=2\line]{Method}{Metrics}} & \multicolumn{5}{c|}{Quadruple Tank} & \multicolumn{5}{c}{Cartpole Balancing} \\
    \cline{2-11}
    & Fail\% & Cost & P-Cost & FLOPs & \#Params & Fail\% & Cost & P-Cost & FLOPs & \#Params \\
    \midrule
    MPC(2)  & 16.59 & 236.1 & 275.7 & 95K$_{+\text{1.15M}}$ & - & 100.0 & 1.36 & 129 & 67K$_{+\text{814K}}$ & - \\
    MPC(16) & 4.27 & 228.3 & 237.2 & 75M$_{+\text{0}}$ & - & 46.86 & 0.34 & 8.39 & 3.9M$_{+\text{47M}}$ & - \\
    MPC-T(2, 1) & 15.75 & 235.0 & 272.2 & 126K$_{+\text{1.12M}}$ & - & 100.0 & 1.39 & 122 & 89K$_{+\text{792K}}$ & - \\
    MPC-T(2, 10) & 10.53 & 226.7 & 250.2 & 220K$_{+\text{1.03M}}$ & - & 100.0 & 1.41 & 122 & 89K$_{+\text{792K}}$ & - \\
    MPC-T(2, 100) & 4.23 & 239.6 & 248.5 & 470K$_{+\text{780K}}$ & - & 100.0 & 1.41 & 122 & 89K$_{+\text{792K}}$ & - \\
    MPC-T(16, 1) & 4.02 & 228.0 & 236.4 & 24M$_{+\text{50M}}$ & - & 33.06 & 0.33 & 4.96 & 3.9M$_{+\text{48M}}$ & - \\
    MPC-T(16, 10) & 3.22 & 224.8 & 231.5 & 26M$_{+\text{48M}}$ & - & 4.74 & 0.30 & 0.79 & 52M$_{+\text{0}}$ & - \\
    MPC-T(16, 100) & 0.37 & 268.5 & 269.2 & 75M$_{+\text{0}}$ & - & 5.01 & 0.34 & 0.87 & 52M$_{+\text{0}}$ & - \\
    RL-MLP(8) & 0.08 & 315.4 & 315.6 & 2K & 952 & 4.52 & 0.94 & 1.41 & 2K & 856 \\
    RL-MLP(16) & 0.00 & 313.3 & 313.3 & 6K & 3.2K & 4.67 & 0.79 & 1.28 & 6K & 3K \\
    RL-MLP(32) & 0.03 & 266.7 & 266.7 & 23K & 11.5K & 4.65 & 0.83 & 1.32 & 23K & 11K \\
    RL-MLP(64) & 0.03 & 291.3 & 291.4 & 87K & 43K & 3.23 & 0.57 & 0.91 & 87K & 43K \\
    LQP(4, 24) & 0.18 & 272.5 & 272.8 & 14K & 354 & 3.49 & 0.76 & 1.12 & 14K & 246 \\
    LQP(8, 48) & 0.22 & 230.7 & 231.1 & 53K & 916 & 3.33 & 0.64 & 0.99 & 53K & 700 \\
    LQP(16, 96) & 0.13 & 227.4 & 227.6 & 208K & 2.7K & 4.11 & 0.44 & 0.87 & 208K & 2.2K \\
    \bottomrule
  \end{tabular}

  \centering
  \begin{tabular}{>{\centering\arraybackslash}m{2.4cm}|ccccc}
    \toprule
    \multirow{2}{*}{\diagbox[dir=NW,innerwidth=2.4cm,height=2\line]{Method}{Metrics}} & \multicolumn{5}{c}{Double Integrator} \\
    \cline{2-6}
    & Fail\% & Cost & P-Cost & FLOPs & \#Params \\
    \midrule
    MPC(2) & 83.80 & 1.78 & 945.06 & 32.8K$_{\text{+218K}}$ & - \\
    MPC(16) & 55.55 & 0.42 & 246.11 & 1.94M$_{\text{+72.6M}}$ & - \\
    MPC-T(2, 1) & 80.10 & 1.16 & 755.63 & 32.8K$_{\text{+218K}}$ & - \\
    MPC-T(2, 10) & 62.52 & 0.69 & 328.06 & 32.8K$_{\text{+312K}}$ & - \\
    MPC-T(2, 100) & 55.43 & 0.82 & 245.31 & 32.8K$_{\text{+468K}}$ & - \\
    MPC-T(16, 1) & 55.55 & 0.42 & 246.11 & 1.94M$_{\text{+72.6M}}$ & - \\
    MPC-T(16, 10) & 55.55 & 0.44 & 246.12 & 1.94M$_{\text{+72.6M}}$ & - \\
    MPC-T(16, 100) & 55.55 & 0.88 & 244.77 & 1.94M$_{\text{+72.6M}}$ & - \\
    RL-MLP(8) & 64.01 & 1.32 & 352.00 & 1.61K & 760 \\
    RL-MLP(16) & 0.0 & 21.00 & 21.00 & 5.78K & 2.8K \\
    RL-MLP(32) & 0.0 & 0.55 & 0.55 & 21.8K & 10.7K \\
    RL-MLP(64) & 0.0 & 1.02 & 1.02 & 84.5K & 41.9K \\
    LQP(4, 24) & 0.0 & 0.56 & 0.56 & 13.8K & 186 \\
    LQP(8, 48) & 0.0 & 0.54 & 0.54 & 53.0K & 580 \\
    LQP(16, 96) & 0.0 & 0.52 & 0.52 & 207K & 2.0K \\
    \bottomrule
  \end{tabular}

  \vspace{0.25em}

  {
  \footnotesize
  Methods: MPC($N$) = naive MPC (Problem~\ref{pb:MPC}) with horizon $N$; MPC-T($N, \rho$) = MPC with horizon $N$ and terminal cost $\rho \| x_N \|^2$; RL-MLP($n$) = reinforcement learning controller with MLP policy with hidden layer sizes $[4n, 2n, n]$; LQP($n_{qp},m_{qp}$) = proposed learned QP controller with problem dimensions $(n_{qp},m_{qp})$. Metrics: see Appendix~\ref{app:metric}.
  }
\end{table}

The benchmarking results for perturbed systems are presented in Table~\ref{tab:extended2}, which extends Table~\ref{tab:robustness} by including more systems as well as baselines. On one hand, under the magnitude of perturbation considered in the experiments, it is shown that the proposed LQP method is more effective and computationally lighter than robust MPC baselines. Although it might be possible to design ad-hoc robust MPC methods tailored for improved performance and efficiency, LQP works off-the-shelf, in the sense that it requires minimal task-specific knowledge, except for a simulator taking the disturbances into account. On the other hand, LQP is even marginally better than MLP controllers, with much fewer learnable policy parameters, which hints at the suitability of the particular parameterized policy class for control tasks.

\begin{table}[!htbp]
  \caption{\small Performance comparison on perturbed systems (extended version).}\label{tab:extended2}
  \small
  \vspace{-10pt}\setlength\tabcolsep{3pt} 
  \renewcommand{\arraystretch}{1.0} 
  \begin{tabular}{>{\centering\arraybackslash}m{2.4cm}|ccccc|ccccc}
    \toprule
    \multirow{2}{*}{\diagbox[dir=NW,innerwidth=2.4cm,height=2\line]{Method}{Metrics}} & \multicolumn{5}{c|}{Quadruple Tank} & \multicolumn{5}{c}{Cartpole Balancing} \\
    \cline{2-11}
    & Fail\% & Cost & P-Cost & Time(s) & \#Params & Fail\% & Cost & P-Cost & Time(s) & \#Params \\
    \midrule
    MPC & 82.6 & 216.8 & 713.4 & 0.25$_{+\text{0.56}}$ & - & 19.1 & 0.08 & 0.57 & 0.06$_{+\text{0}}$ & - \\
    Scenario & 16.4 & 236.9 & 273.2 & 5.21$_{+\text{18}}$ & - & 100.0 & 2.19 & 100.88 & 6.5$_{+\text{1.9}}$ & - \\
    Tube(0.05) & 81.9 & 233.3 & 696.4 & 2.17$_{+\text{39}}$ & - & 100.0 & 2.41 & 112.43 & 3.3$_{+\text{82.5}}$ & - \\
    Tube(0.1) & 82.1 & 218.1 & 675.3 & 2.22$_{+\text{44}}$ & - & 100.0 & 2.41 & 112.43 & 3.3$_{+\text{78.8}}$ & - \\
    Tube(0.2) & 82.5 & 233.3 & 597.9 & 1.98$_{+\text{38}}$ & - & 100.0 & 2.41 & 112.43 & 2.8$_{+\text{59.7}}$ & - \\
    Tube(0.3) & 100.0 & 255.2 & 1322 & 1.60$_{+\text{31}}$ & - & 100.0 & 2.41 & 112.43 & 3.5$_{+\text{89.5}}$ & - \\
    RL-MLP(8) & 2.8 & 249.0 & 254.7 & $2\times 10^{-5}$ & 1.0K & 4.7 & 0.63 & 0.73 & 2K & 856 \\
    RL-MLP(16) & 2.0 & 241.5 & 245.6 & $8\times 10^{-5}$ & 3.2K & 4.8 & 0.40 & 0.50 & 6K & 3K \\
    RL-MLP(32) & 1.5 & 239.2 & 242.3 & $3\times 10^{-4}$ & 11.5K & 3.6 & 0.32 & 0.39 & 23K & 11K \\
    RL-MLP(64) & 1.3 & 238.9 & 241.5 & $1\times 10^{-3}$ & 43K & 3.2 & 0.21 & 0.28 & 87K & 43K \\
    LQP(4, 24) & 2.5 & 256.7 & 261.8 & $2\times 10^{-4}$ & 354 & 3.9 & 0.12 & 0.20 & 14K & 246 \\
    LQP(8, 48) & 1.7 & 240.4 & 243.8 & $7 \times 10^{-4}$ & 916 & 4.3 & 0.11 & 0.20 & 53K & 700 \\
    LQP(16, 96) & 1.4 & 240.6 & 243.4 & $2\times 10^{-3}$ & 2.7K & 4.0 & 0.18 & 0.27 & 208K & 2.2K \\
    \bottomrule
  \end{tabular}

  \centering
  \begin{tabular}{>{\centering\arraybackslash}m{2.4cm}|ccccc}
    \toprule
    \multirow{2}{*}{\diagbox[dir=NW,innerwidth=2.4cm,height=2\line]{Method}{Metrics}} & \multicolumn{5}{c}{Double Integrator} \\
    \cline{2-6}
    & Fail\% & Cost & P-Cost & Time(s) & \#Params \\
    \midrule
    MPC & 36.4 & 0.60 & 114.51 & 0.002$_{\text{+0.02}}$ & - \\
    Scenario & 30.2 & 0.86 & 87.01 & 5.0$_{\text{6.1}}$ & - \\
    Tube(0.05) & 100.0 & 10.58 & 12105 & 0.43$_{\text{+4.66}}$ & - \\
    Tube(0.1) & 100.0 & 10.58 & 12105 & 0.46$_{\text{+3.96}}$ & - \\
    Tube(0.2) & 100.0 & 10.58 & 12105 & 0.46$_{\text{+4.20}}$ & - \\
    Tube(0.3) & 100.0 & 10.58 & 12105 & 0.44$_{\text{+4.38}}$ & - \\
    RL-MLP(8) & 30.2 & 1.18 & 87.33 & $2\times 10^{-5}$ & 760 \\
    RL-MLP(16) & 30.2 & 0.90 & 87.05 & $8\times 10^{-5}$ & 2.8K \\
    RL-MLP(32) & 31.6 & 0.80 & 92.75 & $3 \times 10^{-4}$ & 10.7K \\
    RL-MLP(64) & 31.4 & 0.76 & 87.73 & $1 \times 10^{-3}$ & 41.9K \\
    LQP(4, 24) & 27.5 & 1.03 & 76.58 & $2\times 10^{-4}$ & 186 \\
    LQP(8, 48) & 27.5 & 0.84 & 76.38 & $7\times 10^{-4}$ & 580 \\
    LQP(16, 96) & 27.4 & 0.86 & 76.03 & $2\times 10^{-3}$ & 2.0K \\
    \bottomrule
  \end{tabular}

  \vspace{0.25em}

  {
  \footnotesize
  Methods: MPC = MPC (Problem~\ref{pb:MPC}) with horizon $N=16$ and terminal cost $10 \| x_N \|^2$; Scenario = scenario-based MPC (see Table~\ref{tab:mpc_baselines}); Tube($s$) = tube MPC with tube size $s$ (see Table~\ref{tab:mpc_baselines}); RL-MLP($n$) = reinforcement learning controller with MLP policy with hidden layer sizes $[4n, 2n, n]$; LQP($n_{qp},m_{qp}$) = proposed learned QP controller with problem dimensions $(n_{qp},m_{qp})$. Metrics: see Appendix~\ref{app:metric}.
  }
\end{table}


\subsection{Trajectory-wise Comparisons}

\begin{figure}[!htbp]
  \setlength\figurewidth{0.47\textwidth}
  \setlength\figureheight{0.37\textwidth}
  \subfigure[\small Quadruple tank, MPC vs. LQP]{
\begin{tikzpicture}

    \definecolor{darkgray176}{RGB}{176,176,176}
    \definecolor{steelblue31119180}{RGB}{31,119,180}

    \begin{axis}[
  scaled ticks=false, 
  tick label style={/pgf/number format/fixed, font=\scriptsize}, 
    height=\figureheight,
    tick align=outside,
    tick pos=left,
    width=\figurewidth,
    x grid style={darkgray176},
    xlabel={Ratio of cost (MPC / LQP)},
    xmin=0, xmax=2,
    xtick style={color=black},
    y grid style={darkgray176},
    ylabel={Proportion of trajectories},
    ymin=0, ymax=0.309263895843776,
    ytick style={color=black}
    ]
    \draw[draw=black,fill=steelblue31119180,opacity=0.7] (axis cs:0.00890601052281582,0) rectangle (axis cs:0.0594304960050861,0.0014021031547321);
    \draw[draw=black,fill=steelblue31119180,opacity=0.7] (axis cs:0.0594304960050861,0) rectangle (axis cs:0.109954981487356,0.000801201802704056);
    \draw[draw=black,fill=steelblue31119180,opacity=0.7] (axis cs:0.109954981487356,0) rectangle (axis cs:0.160479466969627,0.000500751126690035);
    \draw[draw=black,fill=steelblue31119180,opacity=0.7] (axis cs:0.160479466969627,0) rectangle (axis cs:0.211003952451897,0.000400600901352028);
    \draw[draw=black,fill=steelblue31119180,opacity=0.7] (axis cs:0.211003952451897,0) rectangle (axis cs:0.261528437934167,0.000400600901352028);
    \draw[draw=black,fill=steelblue31119180,opacity=0.7] (axis cs:0.261528437934167,0) rectangle (axis cs:0.312052923416437,0.000200300450676014);
    \draw[draw=black,fill=steelblue31119180,opacity=0.7] (axis cs:0.312052923416437,0) rectangle (axis cs:0.362577408898708,0.000600901352028042);
    \draw[draw=black,fill=steelblue31119180,opacity=0.7] (axis cs:0.362577408898708,0) rectangle (axis cs:0.413101894380978,0.000200300450676014);
    \draw[draw=black,fill=steelblue31119180,opacity=0.7] (axis cs:0.413101894380978,0) rectangle (axis cs:0.463626379863248,0);
    \draw[draw=black,fill=steelblue31119180,opacity=0.7] (axis cs:0.463626379863248,0) rectangle (axis cs:0.514150865345519,0.000701051577366049);
    \draw[draw=black,fill=steelblue31119180,opacity=0.7] (axis cs:0.514150865345519,0) rectangle (axis cs:0.564675350827789,0.000901352028042063);
    \draw[draw=black,fill=steelblue31119180,opacity=0.7] (axis cs:0.564675350827789,0) rectangle (axis cs:0.615199836310059,0.0014021031547321);
    \draw[draw=black,fill=steelblue31119180,opacity=0.7] (axis cs:0.615199836310059,0) rectangle (axis cs:0.665724321792329,0.00220330495743615);
    \draw[draw=black,fill=steelblue31119180,opacity=0.7] (axis cs:0.665724321792329,0) rectangle (axis cs:0.7162488072746,0.00170255383074612);
    \draw[draw=black,fill=steelblue31119180,opacity=0.7] (axis cs:0.7162488072746,0) rectangle (axis cs:0.76677329275687,0.0042063094641963);
    \draw[draw=black,fill=steelblue31119180,opacity=0.7] (axis cs:0.76677329275687,0) rectangle (axis cs:0.81729777823914,0.00480721081622434);
    \draw[draw=black,fill=steelblue31119180,opacity=0.7] (axis cs:0.81729777823914,0) rectangle (axis cs:0.86782226372141,0.00650976464697046);
    \draw[draw=black,fill=steelblue31119180,opacity=0.7] (axis cs:0.867822263721411,0) rectangle (axis cs:0.918346749203681,0.0102153229844767);
    \draw[draw=black,fill=steelblue31119180,opacity=0.7] (axis cs:0.918346749203681,0) rectangle (axis cs:0.968871234685951,0.0200300450676014);
    \draw[draw=black,fill=steelblue31119180,opacity=0.7] (axis cs:0.968871234685951,0) rectangle (axis cs:1.01939572016822,0.0440660991487227);
    \draw[draw=black,fill=steelblue31119180,opacity=0.7] (axis cs:1.01939572016822,0) rectangle (axis cs:1.06992020565049,0.132699048572859);
    \draw[draw=black,fill=steelblue31119180,opacity=0.7] (axis cs:1.06992020565049,0) rectangle (axis cs:1.12044469113276,0.19328993490236);
    \draw[draw=black,fill=steelblue31119180,opacity=0.7] (axis cs:1.12044469113276,0) rectangle (axis cs:1.17096917661503,0.184276414621939);
    \draw[draw=black,fill=steelblue31119180,opacity=0.7] (axis cs:1.17096917661503,0) rectangle (axis cs:1.2214936620973,0.118778167250875);
    \draw[draw=black,fill=steelblue31119180,opacity=0.7] (axis cs:1.2214936620973,0) rectangle (axis cs:1.27201814757957,0.103855783675512);
    \draw[draw=black,fill=steelblue31119180,opacity=0.7] (axis cs:1.27201814757957,0) rectangle (axis cs:1.32254263306184,0.087130696044065);
    \draw[draw=black,fill=steelblue31119180,opacity=0.7] (axis cs:1.32254263306184,0) rectangle (axis cs:1.37306711854411,0.0480721081622429);
    \draw[draw=black,fill=steelblue31119180,opacity=0.7] (axis cs:1.37306711854411,0) rectangle (axis cs:1.42359160402638,0.0186279419128693);
    \draw[draw=black,fill=steelblue31119180,opacity=0.7] (axis cs:1.42359160402638,0) rectangle (axis cs:1.47411608950865,0.0099148723084627);
    \draw[draw=black,fill=steelblue31119180,opacity=0.7] (axis cs:1.47411608950865,0) rectangle (axis cs:1.52464057499092,0.00210315473209815);
    \path [draw=black, fill=black]
    (axis cs:0.15,0.251276915373068)
    --(axis cs:0.2,0.256276915373068)
    --(axis cs:0.2,0.251776915373068)
    --(axis cs:0.8,0.251776915373068)
    --(axis cs:0.8,0.250776915373068)
    --(axis cs:0.2,0.250776915373068)
    --(axis cs:0.2,0.246276915373068)
    --cycle;
    \path [draw=black, fill=black]
    (axis cs:1.85,0.251276915373068)
    --(axis cs:1.8,0.246276915373068)
    --(axis cs:1.8,0.250776915373068)
    --(axis cs:1.2,0.250776915373068)
    --(axis cs:1.2,0.251776915373068)
    --(axis cs:1.8,0.251776915373068)
    --(axis cs:1.8,0.256276915373068)
    --cycle;
    \addplot [semithick, red, dashed]
    table {%
    1 -6.93889390390723e-18
    1 0.309263895843776
    };
    \draw (axis cs:0.5,0.251276915373068) node[
      anchor=south,
      text=black,
      rotate=0.0
    ]{MPC better\vphantom{Q}};
    \draw (axis cs:1.5,0.251276915373068) node[
      anchor=south,
      text=black,
      rotate=0.0
    ]{LQP better};
    \end{axis}

\end{tikzpicture}
  }
  \subfigure[\small Quadruple tank, MLP vs. LQP]{
\begin{tikzpicture}

    \definecolor{darkgray176}{RGB}{176,176,176}
    \definecolor{steelblue31119180}{RGB}{31,119,180}

    \begin{axis}[
  scaled ticks=false, 
  tick label style={/pgf/number format/fixed, font=\scriptsize}, 
    height=\figureheight,
    tick align=outside,
    tick pos=left,
    width=\figurewidth,
    x grid style={darkgray176},
    xlabel={Log10(Ratio of cost) (MLP / LQP)},
    xmin=-1, xmax=1,
    xtick style={color=black},
    y grid style={darkgray176},
    ylabel={Proportion of trajectories},
    ymin=0, ymax=0.672528793189765,
    ytick style={color=black}
    ]
    \draw[draw=black,fill=steelblue31119180,opacity=0.7] (axis cs:-0.281680935793875,0) rectangle (axis cs:-0.208891451756869,0.000200300450676014);
    \draw[draw=black,fill=steelblue31119180,opacity=0.7] (axis cs:-0.208891451756869,0) rectangle (axis cs:-0.136101967719864,0.000901352028042063);
    \draw[draw=black,fill=steelblue31119180,opacity=0.7] (axis cs:-0.136101967719864,0) rectangle (axis cs:-0.0633124836828588,0.0043064596895343);
    \draw[draw=black,fill=steelblue31119180,opacity=0.7] (axis cs:-0.0633124836828588,0) rectangle (axis cs:0.00947700035414656,0.175663495242869);
    \draw[draw=black,fill=steelblue31119180,opacity=0.7] (axis cs:0.00947700035414656,0) rectangle (axis cs:0.0822664843911519,0.420330495743603);
    \draw[draw=black,fill=steelblue31119180,opacity=0.7] (axis cs:0.0822664843911519,0) rectangle (axis cs:0.155055968428157,0.17305958938408);
    \draw[draw=black,fill=steelblue31119180,opacity=0.7] (axis cs:0.155055968428157,0) rectangle (axis cs:0.227845452465163,0.0847270906359529);
    \draw[draw=black,fill=steelblue31119180,opacity=0.7] (axis cs:0.227845452465163,0) rectangle (axis cs:0.300634936502168,0.0432648973460186);
    \draw[draw=black,fill=steelblue31119180,opacity=0.7] (axis cs:0.300634936502168,0) rectangle (axis cs:0.373424420539173,0.0244366549824736);
    \draw[draw=black,fill=steelblue31119180,opacity=0.7] (axis cs:0.373424420539173,0) rectangle (axis cs:0.446213904576179,0.0156234351527291);
    \draw[draw=black,fill=steelblue31119180,opacity=0.7] (axis cs:0.446213904576179,0) rectangle (axis cs:0.519003388613184,0.0128192288432649);
    \draw[draw=black,fill=steelblue31119180,opacity=0.7] (axis cs:0.519003388613184,0) rectangle (axis cs:0.591792872650189,0.0100150225338007);
    \draw[draw=black,fill=steelblue31119180,opacity=0.7] (axis cs:0.591792872650189,0) rectangle (axis cs:0.664582356687195,0.00721081622433651);
    \draw[draw=black,fill=steelblue31119180,opacity=0.7] (axis cs:0.664582356687195,0) rectangle (axis cs:0.7373718407242,0.00550826239359039);
    \draw[draw=black,fill=steelblue31119180,opacity=0.7] (axis cs:0.7373718407242,0) rectangle (axis cs:0.810161324761205,0.00660991487230847);
    \draw[draw=black,fill=steelblue31119180,opacity=0.7] (axis cs:0.810161324761205,0) rectangle (axis cs:0.882950808798211,0.00400600901352028);
    \draw[draw=black,fill=steelblue31119180,opacity=0.7] (axis cs:0.882950808798211,0) rectangle (axis cs:0.955740292835216,0.00460691036554833);
    \draw[draw=black,fill=steelblue31119180,opacity=0.7] (axis cs:0.955740292835216,0) rectangle (axis cs:1.02852977687222,0.00250375563345018);
    \draw[draw=black,fill=steelblue31119180,opacity=0.7] (axis cs:1.02852977687222,0) rectangle (axis cs:1.10131926090923,0.00170255383074612);
    \draw[draw=black,fill=steelblue31119180,opacity=0.7] (axis cs:1.10131926090923,0) rectangle (axis cs:1.17410874494623,0.00100150225338007);
    \draw[draw=black,fill=steelblue31119180,opacity=0.7] (axis cs:1.17410874494623,0) rectangle (axis cs:1.24689822898324,0.000901352028042063);
    \draw[draw=black,fill=steelblue31119180,opacity=0.7] (axis cs:1.24689822898324,0) rectangle (axis cs:1.31968771302024,0.000200300450676014);
    \draw[draw=black,fill=steelblue31119180,opacity=0.7] (axis cs:1.31968771302024,0) rectangle (axis cs:1.39247719705725,0.000100150225338007);
    \draw[draw=black,fill=steelblue31119180,opacity=0.7] (axis cs:1.39247719705725,0) rectangle (axis cs:1.46526668109425,0.000100150225338007);
    \draw[draw=black,fill=steelblue31119180,opacity=0.7] (axis cs:1.46526668109425,0) rectangle (axis cs:1.53805616513126,0.000100150225338007);
    \draw[draw=black,fill=steelblue31119180,opacity=0.7] (axis cs:1.53805616513126,0) rectangle (axis cs:1.61084564916826,0);
    \draw[draw=black,fill=steelblue31119180,opacity=0.7] (axis cs:1.61084564916826,0) rectangle (axis cs:1.68363513320527,0);
    \draw[draw=black,fill=steelblue31119180,opacity=0.7] (axis cs:1.68363513320527,0) rectangle (axis cs:1.75642461724227,0);
    \draw[draw=black,fill=steelblue31119180,opacity=0.7] (axis cs:1.75642461724227,0) rectangle (axis cs:1.82921410127928,0);
    \draw[draw=black,fill=steelblue31119180,opacity=0.7] (axis cs:1.82921410127928,0) rectangle (axis cs:1.90200358531629,0.000100150225338007);
    \path [draw=black, fill=black]
    (axis cs:-0.85,0.546429644466684)
    --(axis cs:-0.8,0.551429644466684)
    --(axis cs:-0.8,0.546929644466684)
    --(axis cs:-0.2,0.546929644466684)
    --(axis cs:-0.2,0.545929644466684)
    --(axis cs:-0.8,0.545929644466684)
    --(axis cs:-0.8,0.541429644466684)
    --cycle;
    \path [draw=black, fill=black]
    (axis cs:0.85,0.546429644466684)
    --(axis cs:0.8,0.541429644466684)
    --(axis cs:0.8,0.545929644466684)
    --(axis cs:0.2,0.545929644466684)
    --(axis cs:0.2,0.546929644466684)
    --(axis cs:0.8,0.546929644466684)
    --(axis cs:0.8,0.551429644466684)
    --cycle;
    \addplot [semithick, red, dashed]
    table {%
    2.22044604925031e-16 -1.38777878078145e-17
    2.22044604925031e-16 0.672528793189765
    };
    \draw (axis cs:-0.5,0.546429644466684) node[
      anchor=south,
      text=black,
      rotate=0.0
    ]{MLP better\vphantom{Q}};
    \draw (axis cs:0.5,0.546429644466684) node[
      anchor=south,
      text=black,
      rotate=0.0
    ]{LQP better};
    \end{axis}

\end{tikzpicture}
  }
  \\[2ex]
  \subfigure[\small Cartpole balancing, MPC vs. LQP]{
    \hspace{-7pt}
\begin{tikzpicture}

\definecolor{darkgray176}{RGB}{176,176,176}
\definecolor{steelblue31119180}{RGB}{31,119,180}

\begin{axis}[
  scaled ticks=false, 
  tick label style={/pgf/number format/fixed, font=\scriptsize}, 
height=\figureheight,
tick align=outside,
tick pos=left,
width=\figurewidth,
x grid style={darkgray176},
xlabel={Log10(Ratio of cost) (MPC / LQP)},
xmin=-3, xmax=3,
xtick style={color=black},
y grid style={darkgray176},
ylabel={Proportion of trajectories},
ymin=0, ymax=0.151022548505508,
ytick style={color=black}
]
\draw[draw=black,fill=steelblue31119180,opacity=0.7] (axis cs:-2.1995091539005,0) rectangle (axis cs:-2.00618137741639,0.000524383848977452);
\draw[draw=black,fill=steelblue31119180,opacity=0.7] (axis cs:-2.00618137741639,0) rectangle (axis cs:-1.81285360093227,0.00377556371263765);
\draw[draw=black,fill=steelblue31119180,opacity=0.7] (axis cs:-1.81285360093227,0) rectangle (axis cs:-1.61952582444815,0.0130047194546408);
\draw[draw=black,fill=steelblue31119180,opacity=0.7] (axis cs:-1.61952582444815,0) rectangle (axis cs:-1.42619804796404,0.0245411641321448);
\draw[draw=black,fill=steelblue31119180,opacity=0.7] (axis cs:-1.42619804796404,0) rectangle (axis cs:-1.23287027147992,0.0391190351337179);
\draw[draw=black,fill=steelblue31119180,opacity=0.7] (axis cs:-1.23287027147992,0) rectangle (axis cs:-1.0395424949958,0.0463555322496065);
\draw[draw=black,fill=steelblue31119180,opacity=0.7] (axis cs:-1.0395424949958,0) rectangle (axis cs:-0.846214718511684,0.0618772941793388);
\draw[draw=black,fill=steelblue31119180,opacity=0.7] (axis cs:-0.846214718511684,0) rectangle (axis cs:-0.652886942027567,0.0625065547981117);
\draw[draw=black,fill=steelblue31119180,opacity=0.7] (axis cs:-0.652886942027567,0) rectangle (axis cs:-0.45955916554345,0.0786575773466179);
\draw[draw=black,fill=steelblue31119180,opacity=0.7] (axis cs:-0.45955916554345,0) rectangle (axis cs:-0.266231389059333,0.0868379653906665);
\draw[draw=black,fill=steelblue31119180,opacity=0.7] (axis cs:-0.266231389059333,0) rectangle (axis cs:-0.072903612575216,0.0943890928159422);
\draw[draw=black,fill=steelblue31119180,opacity=0.7] (axis cs:-0.072903612575216,0) rectangle (axis cs:0.120424163908901,0.0920818038804413);
\draw[draw=black,fill=steelblue31119180,opacity=0.7] (axis cs:0.120424163908901,0) rectangle (axis cs:0.313751940393018,0.0775039328788675);
\draw[draw=black,fill=steelblue31119180,opacity=0.7] (axis cs:0.313751940393018,0) rectangle (axis cs:0.507079716877135,0.0718405873099108);
\draw[draw=black,fill=steelblue31119180,opacity=0.7] (axis cs:0.507079716877135,0) rectangle (axis cs:0.700407493361252,0.0616675406397478);
\draw[draw=black,fill=steelblue31119180,opacity=0.7] (axis cs:0.700407493361252,0) rectangle (axis cs:0.893735269845369,0.050970110120608);
\draw[draw=black,fill=steelblue31119180,opacity=0.7] (axis cs:0.893735269845369,0) rectangle (axis cs:1.08706304632949,0.0400629260618773);
\draw[draw=black,fill=steelblue31119180,opacity=0.7] (axis cs:1.08706304632949,0) rectangle (axis cs:1.2803908228136,0.0307288935500788);
\draw[draw=black,fill=steelblue31119180,opacity=0.7] (axis cs:1.2803908228136,0) rectangle (axis cs:1.47371859929772,0.0228631358154169);
\draw[draw=black,fill=steelblue31119180,opacity=0.7] (axis cs:1.47371859929772,0) rectangle (axis cs:1.66704637578184,0.0163607760880965);
\draw[draw=black,fill=steelblue31119180,opacity=0.7] (axis cs:1.66704637578184,0) rectangle (axis cs:1.86037415226595,0.00954378605138962);
\draw[draw=black,fill=steelblue31119180,opacity=0.7] (axis cs:1.86037415226595,0) rectangle (axis cs:2.05370192875007,0.00597797587834295);
\draw[draw=black,fill=steelblue31119180,opacity=0.7] (axis cs:2.05370192875007,0) rectangle (axis cs:2.24702970523419,0.00377556371263765);
\draw[draw=black,fill=steelblue31119180,opacity=0.7] (axis cs:2.24702970523419,0) rectangle (axis cs:2.44035748171831,0.00262191924488726);
\draw[draw=black,fill=steelblue31119180,opacity=0.7] (axis cs:2.44035748171831,0) rectangle (axis cs:2.63368525820242,0.000943890928159413);
\draw[draw=black,fill=steelblue31119180,opacity=0.7] (axis cs:2.63368525820242,0) rectangle (axis cs:2.82701303468654,0.000943890928159413);
\draw[draw=black,fill=steelblue31119180,opacity=0.7] (axis cs:2.82701303468654,0) rectangle (axis cs:3.02034081117066,0.000314630309386471);
\draw[draw=black,fill=steelblue31119180,opacity=0.7] (axis cs:3.02034081117066,0) rectangle (axis cs:3.21366858765477,0);
\draw[draw=black,fill=steelblue31119180,opacity=0.7] (axis cs:3.21366858765477,0) rectangle (axis cs:3.40699636413889,0);
\draw[draw=black,fill=steelblue31119180,opacity=0.7] (axis cs:3.40699636413889,0) rectangle (axis cs:3.60032414062301,0.000209753539590981);
\path [draw=black, fill=black]
(axis cs:-2.6,0.122705820660725)
--(axis cs:-2.5,0.125205820660725)
--(axis cs:-2.5,0.123205820660725)
--(axis cs:-0.5,0.123205820660725)
--(axis cs:-0.5,0.122205820660725)
--(axis cs:-2.5,0.122205820660725)
--(axis cs:-2.5,0.120205820660725)
--cycle;
\path [draw=black, fill=black]
(axis cs:2.6,0.122705820660725)
--(axis cs:2.5,0.120205820660725)
--(axis cs:2.5,0.122205820660725)
--(axis cs:0.5,0.122205820660725)
--(axis cs:0.5,0.123205820660725)
--(axis cs:2.5,0.123205820660725)
--(axis cs:2.5,0.125205820660725)
--cycle;
\addplot [semithick, red, dashed]
table {%
0 0
0 0.151022548505508
};
\draw (axis cs:-1.4,0.122705820660725) node[
  anchor=south,
  text=black,
  rotate=0.0
]{MPC better\vphantom{Q}};
\draw (axis cs:1.4,0.122705820660725) node[
  anchor=south,
  text=black,
  rotate=0.0
]{LQP better};
\end{axis}

\end{tikzpicture}
  }
  \subfigure[\small Cartpole balancing, MLP vs. LQP]{
\begin{tikzpicture}

\definecolor{darkgray176}{RGB}{176,176,176}
\definecolor{steelblue31119180}{RGB}{31,119,180}

\begin{axis}[
  scaled ticks=false, 
  tick label style={/pgf/number format/fixed, font=\scriptsize}, 
height=\figureheight,
tick align=outside,
tick pos=left,
width=\figurewidth,
x grid style={darkgray176},
xlabel={Log10(Ratio of cost) (MLP / LQP)},
xmin=-3, xmax=3,
xtick style={color=black},
y grid style={darkgray176},
ylabel={Proportion of trajectories},
ymin=0, ymax=0.158238070267438,
ytick style={color=black}
]
\draw[draw=black,fill=steelblue31119180,opacity=0.7] (axis cs:-1.81788913335914,0) rectangle (axis cs:-1.63121695086939,0.0010487676979549);
\draw[draw=black,fill=steelblue31119180,opacity=0.7] (axis cs:-1.63121695086939,0) rectangle (axis cs:-1.44454476837964,0.00209753539590981);
\draw[draw=black,fill=steelblue31119180,opacity=0.7] (axis cs:-1.44454476837964,0) rectangle (axis cs:-1.25787258588988,0.00702674357629786);
\draw[draw=black,fill=steelblue31119180,opacity=0.7] (axis cs:-1.25787258588988,0) rectangle (axis cs:-1.07120040340013,0.0125852123754588);
\draw[draw=black,fill=steelblue31119180,opacity=0.7] (axis cs:-1.07120040340013,0) rectangle (axis cs:-0.88452822091038,0.0180388044048243);
\draw[draw=black,fill=steelblue31119180,opacity=0.7] (axis cs:-0.88452822091038,0) rectangle (axis cs:-0.697856038420628,0.0293654955427374);
\draw[draw=black,fill=steelblue31119180,opacity=0.7] (axis cs:-0.697856038420628,0) rectangle (axis cs:-0.511183855930876,0.0399580492920818);
\draw[draw=black,fill=steelblue31119180,opacity=0.7] (axis cs:-0.511183855930876,0) rectangle (axis cs:-0.324511673441124,0.0563188253801779);
\draw[draw=black,fill=steelblue31119180,opacity=0.7] (axis cs:-0.324511673441124,0) rectangle (axis cs:-0.137839490951371,0.0713162034609333);
\draw[draw=black,fill=steelblue31119180,opacity=0.7] (axis cs:-0.137839490951371,0) rectangle (axis cs:0.0488326915383808,0.0898793917147359);
\draw[draw=black,fill=steelblue31119180,opacity=0.7] (axis cs:0.0488326915383808,0) rectangle (axis cs:0.235504874028133,0.0935500786575782);
\draw[draw=black,fill=steelblue31119180,opacity=0.7] (axis cs:0.235504874028133,0) rectangle (axis cs:0.422177056517885,0.0988987939171485);
\draw[draw=black,fill=steelblue31119180,opacity=0.7] (axis cs:0.422177056517885,0) rectangle (axis cs:0.608849239007637,0.0896696381751449);
\draw[draw=black,fill=steelblue31119180,opacity=0.7] (axis cs:0.608849239007637,0) rectangle (axis cs:0.795521421497389,0.0819087572102783);
\draw[draw=black,fill=steelblue31119180,opacity=0.7] (axis cs:0.795521421497389,0) rectangle (axis cs:0.982193603987142,0.066386995280545);
\draw[draw=black,fill=steelblue31119180,opacity=0.7] (axis cs:0.982193603987142,0) rectangle (axis cs:1.16886578647689,0.0625065547981117);
\draw[draw=black,fill=steelblue31119180,opacity=0.7] (axis cs:1.16886578647689,0) rectangle (axis cs:1.35553796896665,0.05138961719979);
\draw[draw=black,fill=steelblue31119180,opacity=0.7] (axis cs:1.35553796896665,0) rectangle (axis cs:1.5422101514564,0.0371263765076036);
\draw[draw=black,fill=steelblue31119180,opacity=0.7] (axis cs:1.5422101514564,0) rectangle (axis cs:1.72888233394615,0.0305191400104878);
\draw[draw=black,fill=steelblue31119180,opacity=0.7] (axis cs:1.72888233394615,0) rectangle (axis cs:1.9155545164359,0.0245411641321448);
\draw[draw=black,fill=steelblue31119180,opacity=0.7] (axis cs:1.9155545164359,0) rectangle (axis cs:2.10222669892565,0.0134242265338227);
\draw[draw=black,fill=steelblue31119180,opacity=0.7] (axis cs:2.10222669892565,0) rectangle (axis cs:2.28889888141541,0.00922915574200315);
\draw[draw=black,fill=steelblue31119180,opacity=0.7] (axis cs:2.28889888141541,0) rectangle (axis cs:2.47557106390516,0.00629260618772942);
\draw[draw=black,fill=steelblue31119180,opacity=0.7] (axis cs:2.47557106390516,0) rectangle (axis cs:2.66224324639491,0.00367068694284216);
\draw[draw=black,fill=steelblue31119180,opacity=0.7] (axis cs:2.66224324639491,0) rectangle (axis cs:2.84891542888466,0.00209753539590981);
\draw[draw=black,fill=steelblue31119180,opacity=0.7] (axis cs:2.84891542888466,0) rectangle (axis cs:3.03558761137441,0.000734137388568432);
\draw[draw=black,fill=steelblue31119180,opacity=0.7] (axis cs:3.03558761137441,0) rectangle (axis cs:3.22225979386417,0.000209753539590981);
\draw[draw=black,fill=steelblue31119180,opacity=0.7] (axis cs:3.22225979386417,0) rectangle (axis cs:3.40893197635392,0.00010487676979549);
\draw[draw=black,fill=steelblue31119180,opacity=0.7] (axis cs:3.40893197635392,0) rectangle (axis cs:3.59560415884367,0);
\draw[draw=black,fill=steelblue31119180,opacity=0.7] (axis cs:3.59560415884367,0) rectangle (axis cs:3.78227634133342,0.00010487676979549);
\path [draw=black, fill=black]
(axis cs:-2.6,0.128568432092293)
--(axis cs:-2.5,0.131068432092293)
--(axis cs:-2.5,0.129068432092293)
--(axis cs:-0.5,0.129068432092293)
--(axis cs:-0.5,0.128068432092293)
--(axis cs:-2.5,0.128068432092293)
--(axis cs:-2.5,0.126068432092293)
--cycle;
\path [draw=black, fill=black]
(axis cs:2.6,0.128568432092293)
--(axis cs:2.5,0.126068432092293)
--(axis cs:2.5,0.128068432092293)
--(axis cs:0.5,0.128068432092293)
--(axis cs:0.5,0.129068432092293)
--(axis cs:2.5,0.129068432092293)
--(axis cs:2.5,0.131068432092293)
--cycle;
\addplot [semithick, red, dashed]
table {%
0 0
0 0.158238070267438
};
\draw (axis cs:-1.4,0.128568432092293) node[
  anchor=south,
  text=black,
  rotate=0.0
]{MLP better \vphantom{Q}};
\draw (axis cs:1.4,0.128568432092293) node[
  anchor=south,
  text=black,
  rotate=0.0
]{LQP better};
\end{axis}

\end{tikzpicture}
  }
  \\[2ex]
  \subfigure[\small Double integrator, MPC vs. LQP]{
\begin{tikzpicture}

\definecolor{darkgray176}{RGB}{176,176,176}
\definecolor{steelblue31119180}{RGB}{31,119,180}

\begin{axis}[
  scaled ticks=false, 
  tick label style={/pgf/number format/fixed, font=\scriptsize}, 
height=\figureheight,
tick align=outside,
tick pos=left,
width=\figurewidth,
x grid style={darkgray176},
xlabel={Ratio of average cost (MPC / LQP)},
xmin=0.8, xmax=1.2,
xtick style={color=black},
y grid style={darkgray176},
ylabel={Proportion of trajectories},
ymin=0, ymax=0.257763835984767,
ytick style={color=black}
]
\draw[draw=black,fill=steelblue31119180,opacity=0.7] (axis cs:0.806226488106198,0) rectangle (axis cs:0.81478437941078,0.0369706475464934);
\draw[draw=black,fill=steelblue31119180,opacity=0.7] (axis cs:0.81478437941078,0) rectangle (axis cs:0.823342270715361,0.0259914855478378);
\draw[draw=black,fill=steelblue31119180,opacity=0.7] (axis cs:0.823342270715361,0) rectangle (axis cs:0.831900162019942,0.0197176786914631);
\draw[draw=black,fill=steelblue31119180,opacity=0.7] (axis cs:0.831900162019942,0) rectangle (axis cs:0.840458053324524,0.0161326462021062);
\draw[draw=black,fill=steelblue31119180,opacity=0.7] (axis cs:0.840458053324524,0) rectangle (axis cs:0.849015944629105,0.0179251624467847);
\draw[draw=black,fill=steelblue31119180,opacity=0.7] (axis cs:0.849015944629105,0) rectangle (axis cs:0.857573835933686,0.0244230338337441);
\draw[draw=black,fill=steelblue31119180,opacity=0.7] (axis cs:0.857573835933686,0) rectangle (axis cs:0.866131727238267,0.0266636791395922);
\draw[draw=black,fill=steelblue31119180,opacity=0.7] (axis cs:0.866131727238268,0) rectangle (axis cs:0.874689618542849,0.0378669056688326);
\draw[draw=black,fill=steelblue31119180,opacity=0.7] (axis cs:0.874689618542849,0) rectangle (axis cs:0.88324750984743,0.0345059377100605);
\draw[draw=black,fill=steelblue31119180,opacity=0.7] (axis cs:0.88324750984743,0) rectangle (axis cs:0.891805401152011,0.0273358727313466);
\draw[draw=black,fill=steelblue31119180,opacity=0.7] (axis cs:0.891805401152012,0) rectangle (axis cs:0.900363292456593,0.0338337441183061);
\draw[draw=black,fill=steelblue31119180,opacity=0.7] (axis cs:0.900363292456593,0) rectangle (axis cs:0.908921183761174,0.0564642617073717);
\draw[draw=black,fill=steelblue31119180,opacity=0.7] (axis cs:0.908921183761174,0) rectangle (axis cs:0.917479075065756,0.0434685189334528);
\draw[draw=black,fill=steelblue31119180,opacity=0.7] (axis cs:0.917479075065756,0) rectangle (axis cs:0.926036966370337,0.0750616177459108);
\draw[draw=black,fill=steelblue31119180,opacity=0.7] (axis cs:0.926036966370337,0) rectangle (axis cs:0.934594857674918,0.0663231010531033);
\draw[draw=black,fill=steelblue31119180,opacity=0.7] (axis cs:0.934594857674918,0) rectangle (axis cs:0.943152748979499,0.0790947792964373);
\draw[draw=black,fill=steelblue31119180,opacity=0.7] (axis cs:0.943152748979499,0) rectangle (axis cs:0.951710640284081,0.161102397490479);
\draw[draw=black,fill=steelblue31119180,opacity=0.7] (axis cs:0.951710640284081,0) rectangle (axis cs:0.960268531588662,0.102173425946673);
\draw[draw=black,fill=steelblue31119180,opacity=0.7] (axis cs:0.960268531588662,0) rectangle (axis cs:0.968826422893243,0.0300246470983643);
\draw[draw=black,fill=steelblue31119180,opacity=0.7] (axis cs:0.968826422893243,0) rectangle (axis cs:0.977384314197825,0.0602733587273134);
\draw[draw=black,fill=steelblue31119180,opacity=0.7] (axis cs:0.977384314197825,0) rectangle (axis cs:0.985942205502406,0.0219583239973112);
\draw[draw=black,fill=steelblue31119180,opacity=0.7] (axis cs:0.985942205502406,0) rectangle (axis cs:0.994500096806987,0.000896258122339234);
\draw[draw=black,fill=steelblue31119180,opacity=0.7] (axis cs:0.994500096806987,0) rectangle (axis cs:1.00305798811157,0.00134438718350885);
\draw[draw=black,fill=steelblue31119180,opacity=0.7] (axis cs:1.00305798811157,0) rectangle (axis cs:1.01161587941615,0);
\draw[draw=black,fill=steelblue31119180,opacity=0.7] (axis cs:1.01161587941615,0) rectangle (axis cs:1.02017377072073,0);
\draw[draw=black,fill=steelblue31119180,opacity=0.7] (axis cs:1.02017377072073,0) rectangle (axis cs:1.02873166202531,0);
\draw[draw=black,fill=steelblue31119180,opacity=0.7] (axis cs:1.02873166202531,0) rectangle (axis cs:1.03728955332989,0.000224064530584808);
\draw[draw=black,fill=steelblue31119180,opacity=0.7] (axis cs:1.03728955332989,0) rectangle (axis cs:1.04584744463448,0);
\draw[draw=black,fill=steelblue31119180,opacity=0.7] (axis cs:1.04584744463448,0) rectangle (axis cs:1.05440533593906,0);
\draw[draw=black,fill=steelblue31119180,opacity=0.7] (axis cs:1.05440533593906,0) rectangle (axis cs:1.06296322724364,0.000224064530584808);
\path [draw=black, fill=black]
(axis cs:0.81,0.209433116737623)
--(axis cs:0.83,0.211933116737623)
--(axis cs:0.83,0.209933116737623)
--(axis cs:0.98,0.209933116737623)
--(axis cs:0.98,0.208933116737623)
--(axis cs:0.83,0.208933116737623)
--(axis cs:0.83,0.206933116737623)
--cycle;
\path [draw=black, fill=black]
(axis cs:1.19,0.209433116737623)
--(axis cs:1.17,0.206933116737623)
--(axis cs:1.17,0.208933116737623)
--(axis cs:1.02,0.208933116737623)
--(axis cs:1.02,0.209933116737623)
--(axis cs:1.17,0.209933116737623)
--(axis cs:1.17,0.211933116737623)
--cycle;
\addplot [semithick, red, dashed]
table {%
1 0
1 0.257763835984767
};
\draw (axis cs:0.9,0.209433116737623) node[
  anchor=south,
  text=black,
  rotate=0.0
]{MPC better\vphantom{Q}};
\draw (axis cs:1.1,0.209433116737623) node[
  anchor=south,
  text=black,
  rotate=0.0
]{LQP better};
\end{axis}

\end{tikzpicture}
  }
  \subfigure[\small Double integrator, MLP vs. LQP]{
    \hspace{-6pt}
\begin{tikzpicture}

\definecolor{darkgray176}{RGB}{176,176,176}
\definecolor{steelblue31119180}{RGB}{31,119,180}

\begin{axis}[
  scaled ticks=false, 
  tick label style={/pgf/number format/fixed, font=\scriptsize}, 
height=\figureheight,
tick align=outside,
tick pos=left,
width=\figurewidth,
x grid style={darkgray176},
xlabel={Log10(Ratio of average cost) (MLP / LQP)},
xmin=-1, xmax=1,
xtick style={color=black},
y grid style={darkgray176},
ylabel={Proportion of trajectories},
ymin=0, ymax=1.1620799999999,
ytick style={color=black}
]
\draw[draw=black,fill=steelblue31119180,opacity=0.7] (axis cs:-0.0679120960162165,0) rectangle (axis cs:0.00563102867802867,0.726299999999936);
\draw[draw=black,fill=steelblue31119180,opacity=0.7] (axis cs:0.00563102867802867,0) rectangle (axis cs:0.0791741533722739,0.118000000000002);
\draw[draw=black,fill=steelblue31119180,opacity=0.7] (axis cs:0.0791741533722738,0) rectangle (axis cs:0.152717278066519,0.0407000000000001);
\draw[draw=black,fill=steelblue31119180,opacity=0.7] (axis cs:0.152717278066519,0) rectangle (axis cs:0.226260402760764,0.0249999999999999);
\draw[draw=black,fill=steelblue31119180,opacity=0.7] (axis cs:0.226260402760764,0) rectangle (axis cs:0.299803527455009,0.0195999999999999);
\draw[draw=black,fill=steelblue31119180,opacity=0.7] (axis cs:0.299803527455009,0) rectangle (axis cs:0.373346652149255,0.0148);
\draw[draw=black,fill=steelblue31119180,opacity=0.7] (axis cs:0.373346652149255,0) rectangle (axis cs:0.4468897768435,0.0145);
\draw[draw=black,fill=steelblue31119180,opacity=0.7] (axis cs:0.4468897768435,0) rectangle (axis cs:0.520432901537745,0.0133);
\draw[draw=black,fill=steelblue31119180,opacity=0.7] (axis cs:0.520432901537745,0) rectangle (axis cs:0.59397602623199,0.012);
\draw[draw=black,fill=steelblue31119180,opacity=0.7] (axis cs:0.59397602623199,0) rectangle (axis cs:0.667519150926235,0.00750000000000001);
\draw[draw=black,fill=steelblue31119180,opacity=0.7] (axis cs:0.667519150926235,0) rectangle (axis cs:0.741062275620481,0.0066);
\draw[draw=black,fill=steelblue31119180,opacity=0.7] (axis cs:0.741062275620481,0) rectangle (axis cs:0.814605400314726,0.0009);
\draw[draw=black,fill=steelblue31119180,opacity=0.7] (axis cs:0.814605400314726,0) rectangle (axis cs:0.888148525008971,0.0003);
\draw[draw=black,fill=steelblue31119180,opacity=0.7] (axis cs:0.888148525008971,0) rectangle (axis cs:0.961691649703216,0.0002);
\draw[draw=black,fill=steelblue31119180,opacity=0.7] (axis cs:0.961691649703216,0) rectangle (axis cs:1.03523477439746,0);
\draw[draw=black,fill=steelblue31119180,opacity=0.7] (axis cs:1.03523477439746,0) rectangle (axis cs:1.10877789909171,0);
\draw[draw=black,fill=steelblue31119180,opacity=0.7] (axis cs:1.10877789909171,0) rectangle (axis cs:1.18232102378595,0);
\draw[draw=black,fill=steelblue31119180,opacity=0.7] (axis cs:1.18232102378595,0) rectangle (axis cs:1.2558641484802,0);
\draw[draw=black,fill=steelblue31119180,opacity=0.7] (axis cs:1.2558641484802,0) rectangle (axis cs:1.32940727317444,0.0001);
\draw[draw=black,fill=steelblue31119180,opacity=0.7] (axis cs:1.32940727317444,0) rectangle (axis cs:1.40295039786869,0);
\draw[draw=black,fill=steelblue31119180,opacity=0.7] (axis cs:1.40295039786869,0) rectangle (axis cs:1.47649352256293,0);
\draw[draw=black,fill=steelblue31119180,opacity=0.7] (axis cs:1.47649352256293,0) rectangle (axis cs:1.55003664725718,0);
\draw[draw=black,fill=steelblue31119180,opacity=0.7] (axis cs:1.55003664725718,0) rectangle (axis cs:1.62357977195142,0);
\draw[draw=black,fill=steelblue31119180,opacity=0.7] (axis cs:1.62357977195142,0) rectangle (axis cs:1.69712289664567,0);
\draw[draw=black,fill=steelblue31119180,opacity=0.7] (axis cs:1.69712289664567,0) rectangle (axis cs:1.77066602133991,0);
\draw[draw=black,fill=steelblue31119180,opacity=0.7] (axis cs:1.77066602133991,0) rectangle (axis cs:1.84420914603416,0.0001);
\draw[draw=black,fill=steelblue31119180,opacity=0.7] (axis cs:1.84420914603416,0) rectangle (axis cs:1.9177522707284,0);
\draw[draw=black,fill=steelblue31119180,opacity=0.7] (axis cs:1.9177522707284,0) rectangle (axis cs:1.99129539542265,0);
\draw[draw=black,fill=steelblue31119180,opacity=0.7] (axis cs:1.99129539542265,0) rectangle (axis cs:2.06483852011689,0);
\draw[draw=black,fill=steelblue31119180,opacity=0.7] (axis cs:2.06483852011689,0) rectangle (axis cs:2.13838164481114,0.0001);
\path [draw=black, fill=black]
(axis cs:-0.85,0.944189999999917)
--(axis cs:-0.8,0.949189999999917)
--(axis cs:-0.8,0.944689999999917)
--(axis cs:-0.2,0.944689999999917)
--(axis cs:-0.2,0.943689999999917)
--(axis cs:-0.8,0.943689999999917)
--(axis cs:-0.8,0.939189999999917)
--cycle;
\path [draw=black, fill=black]
(axis cs:0.85,0.944189999999917)
--(axis cs:0.8,0.939189999999917)
--(axis cs:0.8,0.943689999999917)
--(axis cs:0.2,0.943689999999917)
--(axis cs:0.2,0.944689999999917)
--(axis cs:0.8,0.944689999999917)
--(axis cs:0.8,0.949189999999917)
--cycle;
\addplot [semithick, red, dashed]
table {%
2.22044604925031e-16 0
2.22044604925031e-16 1.1620799999999
};
\draw (axis cs:-0.5,0.944189999999917) node[
  anchor=south,
  text=black,
  rotate=0.0
]{MLP better\vphantom{Q}};
\draw (axis cs:0.5,0.944189999999917) node[
  anchor=south,
  text=black,
  rotate=0.0
]{LQP better};
\end{axis}

\end{tikzpicture}
  }
  \caption{\small Histogram of the ratio of MPC/MLP cost to LQP cost. For each task, each pair of controllers are run on 10,000 different trials. In each trial, the same initialization is used for both controllers. The ratios of cumulative costs among trials completed by both controllers are plotted. In some subfigures, the x-axis is in log scale for the ease of observation.}
  \label{fig:cost_ratio}
\end{figure}
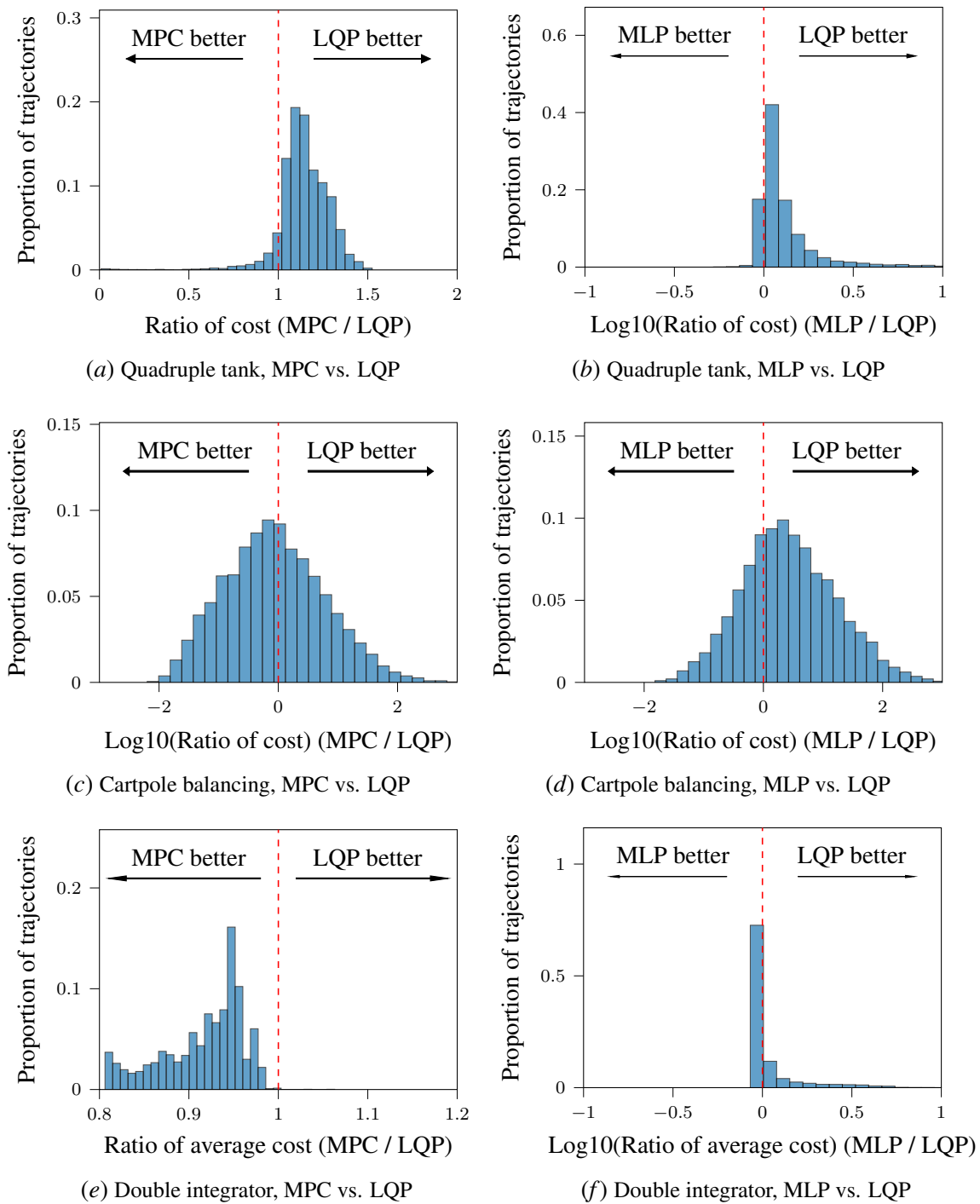

In this subsection, we provide trajectory-wise comparisons of the proposed LQP controllers with MPC and MLP controllers. For each task, we use the nominal system, and randomize the initialization for 10,000 times. Under each initialization, we run the controllers LQP(16, 96), MPC(16, 10) and RL-MLP(32) in Table~\ref{tab:extended1}.
We do pairwise comparisons of (MPC, LQP) and (MLP, LQP) in terms of the cumulative cost along the trajectories: for each initialization where both controllers can run until the maximum number of steps without violating the constraints, we record the ratios of cumulative costs along the trajectories, i.e., (Cost of MPC / Cost of LQP) or (Cost of MLP / Cost of LQP). The histograms of these ratios are plotted in Figure~\ref{fig:cost_ratio}. The following observations can be made from the histograms:
\begin{itemize}
  \item LQP is overall competitive with MPC: it slightly outperforms MPC in the quadruple tank task, is approximately on par with MPC in the cartpole balancing task, and is slightly worse in the double integrator task. The last is reasonable, since it is evident from Table~\ref{tab:extended1} that MPC fails a significant proportion of the trials while LQP does not, and the cost of the failed trials is not included in the histogram. That said, the cost ratio is still close to 1, indicating the competitiveness of LQP. The slight suboptimality of LQP can be attributed to the RL algorithm used for training, the improvement of which is a potential future work.
  \item LQP can act as a drop-in replacement for MLP in control tasks: LQP can slightly outperform MLP on average, and furthermore, in all of the three tasks considered, the cost ratios of (MLP/LQP) are \emph{heavy-tailed} and inclined towards the right, showing that LQP can effectively mitigate the extreme cases where MLP performs poorly. This is an empirical counterpart of the verifiability of LQP brought by its structure: it can potentially improve the reliability of the learned controllers.
  \item LQP can display distinct behavior compared to MPC and MLP: it is evident that LQP neither dominates over MPC or MLP, nor vice versa. Especially, in the cartpole task, the cost of LQP and be either 100 times higher than that of MPC/MLP, or 100 times lower. This is an interesting phenomenon indicating that LQP can potentially provide a different perspective on the control problem, though further investigation is needed to understand the benefits and/or drawbacks of this distinct behavior.
\end{itemize}

\end{document}